\documentclass[12pt]{article}
\usepackage{amsfonts,graphicx,amsmath,amsthm,amssymb,epsfig}
\usepackage{float}
\allowdisplaybreaks
\begin{document}

\title{\bf Complexity of Dynamical Dissipative Cylindrical System in Non-minimally Coupled Theory}
\author{M. Sharif \thanks{msharif.math@pu.edu.pk} and T. Naseer \thanks{tayyabnaseer48@yahoo.com}\\
Department of Mathematics, University of the Punjab,\\
Quaid-i-Azam Campus, Lahore-54590, Pakistan.}

\date{}
\maketitle

\begin{abstract}
This paper aims to formulate certain scalar factors associated with
matter variables for self-gravitating non-static cylindrical
geometry by considering a standard model
$\mathcal{R}+\zeta\mathcal{Q}$ of
$f(\mathcal{R},\mathcal{T},\mathcal{Q})$ gravity, where
$\mathcal{Q}=\mathcal{R}_{\varphi\vartheta}\mathcal{T}^{\varphi\vartheta}$
and $\zeta$ is the arbitrary coupling parameter. We split the
Riemann tensor orthogonally to calculate four scalars and deduce
$\mathcal{Y}_{TF}$ as complexity factor for the fluid configuration.
This scalar incorporates the influence of inhomogeneous energy
density, heat flux and pressure anisotropy along with correction
terms of the modified gravity. We discuss the dynamics of cylinder
by considering two simplest modes of structural evolution. We then
take $\mathcal{Y}_{TF}=0$ with homologous condition to determine the
solution for dissipative as well as non-dissipative scenarios.
Finally, we discuss the criterion under which the complexity-free
condition shows stable behavior throughout the evolution. It is
concluded that complex functional of this theory results in a more
complex structure.
\end{abstract}

\section{Introduction}

The universe includes massive and extremely dense structures ranging
from small planets to galaxies and their clusters. They provide
information about cosmic evolution and can be considered as
foundation of the cosmological study. In this perspective, Einstein
proposed general theory of relativity ($\mathbb{GR}$) which helps
astrophysicists to deal with the challenging issues. Later, multiple
extensions to $\mathbb{GR}$ were proposed to discuss different
issues that $\mathbb{GR}$ cannot deal properly like accelerated
expansion of the universe. It has been pointed out through various
observations that the rapid cosmic expansion is triggered due to the
presence of an obscure force on large scales, named as dark energy
which has immense repulsive nature. Another unknown element of the
cosmos is dark matter whose existence has been verified by several
experiments, i.e., dynamics of galactic clusters and rotation curves
of spiral galaxies. General relativity was firstly extended to
$f(\mathcal{R})$ theory \cite{1} by replacing the Ricci scalar with
its generic function $f(\mathcal{R})$ in the Einstein-Hilbert action
to study the cosmic features at large scale. Various authors
\cite{2}-\cite{2d} have explored physical feasibility of compact
structures with the help of multiple approaches in this framework.

The appealing nature of the cosmos prompted Bertolami et al.
\cite{10} to discover some more fascinating results about its
composition. Firstly, they introduced the idea of coupling between
matter and geometry in $f(\mathcal{R})$ framework by considering the
Lagrangian as a function of $\mathcal{R}$ and $\mathbb{L}_{m}$.
Based on such coupling, several extensions of $\mathbb{GR}$ were
proposed to analyze the physical acceptance of different modified
gravitational models. Harko et al. \cite{20} introduced
$f(\mathcal{R},\mathcal{T})$ theory, in which $\mathcal{T}$ is trace
of the energy-momentum tensor $(\mathbb{EMT})$. The arbitrarily
coupled theories results in the non-vanishing divergence of the
corresponding $\mathbb{EMT}$ opposing $\mathbb{GR}$ as well as
$f(\mathcal{R})$ gravity. This coupling theory is extensively
studied in the analysis of self-gravitating celestial structures and
several remarkable results have been found \cite{21}-\cite{21f}.
Zubair et al. \cite{21g} analyzed dynamical cylinder whose interior
is filled with anisotropic fluid in this theory and derived
evolution equation by applying perturbation approach. They also
investigated instability range for the considered setup in different
regimes through an adiabatic index. The coupling effects of this
theory on geometrical structure disappear when the configuration
involves traceless $\mathbb{EMT}$, i.e., $\mathcal{T}=0$.

Haghani et al. \cite{22} proposed the most general form of the
Lagrangian whose functional depends on $\mathcal{R},~\mathcal{T}$
and $\mathcal{Q}$. This theory helps to understand the cosmic
inflationary era properly. The three different models of this theory
such as
$\mathcal{R}+\lambda\mathcal{Q},~\mathcal{R}(1+\lambda\mathcal{Q})$
and $\mathcal{R}+\zeta\sqrt {|\mathcal{T}|}+\lambda\mathcal{Q}$
along with their cosmological applications have been studied. Sharif
and Zubair \cite{22a} adopted the first two models with different
choices of matter Lagrangian as $\mathbb{L}_m=\rho,~-p$ to discuss
thermodynamical laws of black hole and determined their respective
viability constraints. They also discussed validity of energy bounds
for both modified models to obtain acceptable values of $\lambda$
and concluded that its negative values do not fulfill weak energy
conditions in this framework \cite{22b}.

Odintsov and S\'{a}ez-G\'{o}mez \cite{23} constructed solutions to
many cosmological models and verified that the $\Lambda$CDM model is
supported by $f(\mathcal{R},\mathcal{T},\mathcal{Q})$ gravity. They
also found that this theory may permit to produce pure de Sitter
universe. Baffou et al. \cite{25} considered two mathematical models
and solved the corresponding Friedmann equations as well as
perturbation functions through numerical techniques to explore their
stability. Sharif and Waseem \cite{25a} adopted matter Lagrangian as
$\mathbb{L}_m=-p_r,~-p_t$ to solve the modified field equations for
anisotropic configuration and found that $\mathbb{L}_m=-p_r$
produces more stable structures. Yousaf et al. \cite{26}-\cite{26c}
considered the Riemann curvature tensor and split it orthogonally to
calculate some scalar factors with/without the presence of
electromagnetic field for spherical spacetime. The same authors
\cite{26d,26e} also utilized this technique for static cylinder and
concluded $\mathcal{Y}_{TF}$ as complexity factor that helps to
understand the evolution of self-gravitating structures. We employed
different approaches to obtain charged/uncharged anisotropic
solutions to the field equations and found them physically
acceptable \cite{27}-\cite{27c}.

The observable portion of the universe is mainly made up of massive
and extremely dense bodies such as stars and galaxies, etc. The
interior of these self-gravitating bodies incorporate certain
physical quantities (energy density, pressure, heat flow) that
increase their complexity. Thus, some mathematical notion is
required in terms of these physical factors to define structural
complexity which has been done through numerous attempts in multiple
scientific disciplines. L{\'o}pez-Ruiz et al. \cite{28} gave the
first ever idea of complexity by defining it in terms of information
and entropy. Initially, two simplest physical structures (perfect
crystal and ideal gas) were considered to calculate their
complexity. The arrangement of molecules in former structure is
thoroughly symmetric and has no entropy while it is maximum in later
system due to randomly dispersed particles. Further, they provide
less and maximum information according to their systematic
distribution. However, there is no complexity in both structures.

Later, another definition to analyze complexity was developed in
terms of disequilibrium under which the complexity of ideal gas and
perfect crystal has been found to be zero \cite{29,29a}. The
complexity of different physical systems was calculated by replacing
the probability distribution with energy density \cite{30,30a}, yet
this definition was insufficient due to non-involvement of many
other state variables (heat flux, temperature and pressure, etc.).
Recently, Herrera \cite{31} proposed a new concept by redefining
complexity in terms of pressure anisotropy and inhomogeneous energy
density for a static sphere. Some scalars have been deduced from the
orthogonal splitting of the Riemann tensor and named one of them as
the complexity factor that involves all aforesaid parameters. Sharif
and Butt \cite{32,32a} extended this concept for the case of charged
spherical as well as uncharged cylindrical matter sources. Herrera
et al. \cite{33} studied the complexity for a non-static system in
the presence of dissipative flux and discussed some evolutionary
patterns. They also generalized this definition to the axially
symmetric geometry \cite{34}. Sharif and Majid \cite{35}-\cite{35c}
extended this work in Brans-Dicke scenario and found several
solutions corresponding to complicated systems. Zubair and Azmat
\cite{36a,36b} studied non-static spherically as well as
cylindrically symmetric dynamical objects through the complexity
factor in the background of $f(\mathcal{R},\mathcal{T})$ gravity.
They also discussed different patterns of evolution and obtained
stability of vanishing of the complexity factor. The complexity for
dynamical self-gravitating objects has also been analyzed in
$f(\mathcal{G},\mathcal{T})$ theory, where $\mathcal{G}$ is the
Gauss-Bonnet invariant \cite{36d,36e}.

This paper investigates physical quantities that produce complexity
interior to the dynamical cylinder for the
$\mathcal{R}+\zeta\mathcal{R}_{\varphi\vartheta}\mathcal{T}^{\varphi\vartheta}$
gravity model. Following is the structure of this paper. In section
\textbf{2}, we define some basic definitions and compute non-null
components of the modified field equations as well as Bianchi
identities. Section \textbf{3} provides orthogonal splitting of the
Riemann tensor that yields four scalar functions. Two evolutionary
modes, namely homologous and homogeneous expansion are discussed in
section \textbf{4}. In section \textbf{5}, some kinematical as well
as dynamical quantities are derived to obtain possible solutions in
dissipative/non-dissipative modes. Section \textbf{6} offers
conditions under which the system is departed from zero complexity
condition. We summarize all our findings in section \textbf{7}.

\section{The $f(\mathcal{R},\mathcal{T},\mathcal{R}_{\varphi\vartheta}\mathcal{T}^{\varphi\vartheta})$ Gravity}

The Einstein-Hilbert action ($\kappa=8\pi$) of this theory is
modified by replacing the Ricci scalar $\mathcal{R}$ with a generic
functional of $\mathcal{R},~\mathcal{T}$ and
$\mathcal{R}_{\varphi\vartheta}\mathcal{T}^{\varphi\vartheta}$ as
\cite{23}
\begin{equation}\label{g1}
\mathbb{S}_{f(\mathcal{R},\mathcal{T},\mathcal{R}_{\varphi\vartheta}\mathcal{T}^{\varphi\vartheta})}=\int\sqrt{-g}
\left\{\frac{f(\mathcal{R},\mathcal{T},\mathcal{R}_{\varphi\vartheta}\mathcal{T}^{\varphi\vartheta})}{16\pi}
+\mathbb{L}_{m}\right\}d^{4}x,
\end{equation}
where the matter Lagrangian $\mathbb{L}_{m}$ involves the strong
non-minimal coupling of fluid distribution and geometry. After
executing the variational principle on action \eqref{g1}, the field
equations take the form
\begin{equation}\label{g2}
\mathcal{G}_{\varphi\vartheta}=8\pi\mathcal{T}_{\varphi\vartheta}^{(\mathrm{EFF})}=8\pi\bigg\{\frac{\mathcal{T}_{\varphi\vartheta}}
{f_{\mathcal{R}}-\mathbb{L}_{m}f_{\mathcal{Q}}}+\mathcal{T}_{\varphi\vartheta}^{(\mathcal{C})}\bigg\},
\end{equation}
which explains the geometry in terms of spacetime curvature. Here,
$\mathcal{G}_{\varphi\vartheta}$ and
$\mathcal{T}_{\varphi\vartheta}^{(\mathrm{EFF})}$ are termed as the
Einstein tensor and the modified $\mathbb{EMT}$ which interlinks
geometrical quantities with state variables and their derivatives.
In this regard, the sector
$\mathcal{T}_{\varphi\vartheta}^{(\mathcal{C})}$ becomes
\begin{eqnarray}\nonumber
\mathcal{T}_{\varphi\vartheta}^{(\mathcal{C})}&=&-\frac{1}{8\pi\bigg(\mathbb{L}_{m}f_{\mathcal{Q}}-f_{\mathcal{R}}\bigg)}
\left[\left(f_{\mathcal{T}}+\frac{1}{2}\mathcal{R}f_{\mathcal{Q}}\right)\mathcal{T}_{\varphi\vartheta}
+\left\{\frac{\mathcal{R}}{2}(\frac{f}{\mathcal{R}}-f_{\mathcal{R}})-\mathbb{L}_{m}f_{\mathcal{T}}\right.\right.\\\nonumber
&-&\left.\frac{1}{2}\nabla_{\varrho}\nabla_{\omega}(f_{\mathcal{Q}}\mathcal{T}^{\varrho\omega})\right\}g_{\varphi\vartheta}
-\frac{1}{2}\Box(f_{\mathcal{Q}}\mathcal{T}_{\varphi\vartheta})-(g_{\varphi\vartheta}\Box-
\nabla_{\varphi}\nabla_{\vartheta})f_{\mathcal{R}}\\\label{g4}
&-&2f_{\mathcal{Q}}\mathcal{R}_{\varrho(\varphi}\mathcal{T}_{\vartheta)}^{\varrho}
+\nabla_{\varrho}\nabla_{(\varphi}[\mathcal{T}_{\vartheta)}^{\varrho}
f_{\mathcal{Q}}]+2(f_{\mathcal{Q}}\mathcal{R}^{\varrho\omega}+\left.f_{\mathcal{T}}g^{\varrho\omega})\frac{\partial^2\mathbb{L}_{m}}
{\partial g^{\varphi\vartheta}\partial g^{\varrho\omega}}\right].
\end{eqnarray}
The partial derivatives $f_{\mathcal{R}},~f_{\mathcal{T}}$ and
$f_{\mathcal{Q}}$ in the above equation demonstrate $\frac{\partial
f(\mathcal{R},\mathcal{T},\mathcal{Q})}{\partial \mathcal{R}}$,
$\frac{\partial f(\mathcal{R},\mathcal{T},\mathcal{Q})}{\partial
\mathcal{T}}$ and $\frac{\partial
f(\mathcal{R},\mathcal{T},\mathcal{Q})}{\partial \mathcal{Q}}$,
respectively. Also, $\nabla_\varrho$ means covariant derivative and
$\Box\equiv
\frac{1}{\sqrt{-g}}\partial_\varphi\big(\sqrt{-g}g^{\varphi\vartheta}\partial_{\vartheta}\big)$
is the D'Alambert operator.

We consider $\mathbb{L}_{m}=-\mu$, where $\mu$ represents the energy
density of inner configuration, thus the last term on the right side
of Eq.\eqref{g4} involving matter Lagrangian leads to
$\frac{\partial^2\mathbb{L}_{m}} {\partial
g^{\varphi\vartheta}\partial g^{\varrho\omega}}=0$ \cite{22}. The
$\mathbb{EMT}$ provides internal anisotropic configuration in the
presence of heat dissipation as
\begin{equation}\label{g5}
\mathcal{T}_{\varphi\vartheta}=\mu \mathcal{K}_{\varphi}
\mathcal{K}_{\vartheta}+P
h_{\varphi\vartheta}+\Pi_{\varphi\vartheta}+\varsigma\big(\mathcal{K}_\varphi\mathcal{W}_\vartheta
+\mathcal{W}_\varphi\mathcal{K}_\vartheta\big).
\end{equation}
It should be noted that the pressure generally exists in three
different directions for cylindrically symmetric anisotropic
structure, but the $\mathbb{EMT}$ \eqref{g5} is not the most general
form of the fluid distribution. Here, the four-vector is denoted by
$\mathcal{W}_{\varphi}$ and four-velocity is indicated by
$\mathcal{K}_\varphi$. Also, $h_{\varphi\vartheta}$ and $\varsigma$
represent the projection tenor and heat flux, respectively. These
quantities satisfy the following relations
\begin{align}\label{g5a}
\mathcal{K}_{\varphi} \mathcal{K}^{\varphi}=-1, \quad
\mathcal{K}_{\varphi} \mathcal{W}^{\varphi}=0, \quad
\varsigma_{\varphi} \mathcal{K}^{\varphi}=0, \quad
\mathcal{W}_{\varphi} \mathcal{W}^{\varphi}=1.
\end{align}
The remaining quantities have the form as
\begin{align}\label{g5b}
\Pi_{\varphi\vartheta}&=\Pi\bigg(\mathcal{W}_{\varphi}\mathcal{W}_{\vartheta}-\frac{h_{\varphi\vartheta}}{3}\bigg),
\quad \Pi=P_r-P_\bot,\\\label{g5c}  P&=\frac{P_r+2P_\bot}{3}, \quad
h_{\varphi\vartheta}=g_{\varphi\vartheta}+\mathcal{K}_{\varphi}
\mathcal{K}_{\vartheta}.
\end{align}
Due to the involvement of matter-geometry coupling, this theory
violates the equivalence principle which produces non-null
divergence of the $\mathbb{EMT}$ \eqref{g4} (i.e., $\nabla_\varphi
\mathcal{T}^{\varphi\vartheta}\neq 0$) dissimilar to other extended
theories \cite{39,40}. This produces an additional force which
causes test particles to move in non-geodesic path in their
gravitational field, hence
\begin{align}\nonumber
\nabla^\varphi
\mathcal{T}_{\varphi\vartheta}&=\frac{2}{2f_\mathcal{T}+\mathcal{R}f_\mathcal{Q}+16\pi}\bigg[\nabla_\varphi
\big(f_\mathcal{Q}\mathcal{R}^{\varrho\varphi}\mathcal{T}_{\varrho\vartheta}\big)-\mathcal{G}_{\varphi\vartheta}\nabla^\varphi
\big(f_\mathcal{Q}\mathbb{L}_m\big)+\nabla_\vartheta\big(\mathbb{L}_mf_\mathcal{T}\big)\\\label{g4a}
&-\frac{1}{2}\nabla_\vartheta\mathcal{T}^{\varrho\omega}\big(f_\mathcal{T}g_{\varrho\omega}+f_\mathcal{Q}\mathcal{R}_{\varrho\omega}\big)
-\frac{1}{2}\big\{\nabla^{\varphi}(\mathcal{R}f_{\mathcal{Q}})
+2\nabla^{\varphi}f_{\mathcal{T}}\big\}\mathcal{T}_{\varphi\vartheta}\bigg].
\end{align}
The trace of the field equations turns out to be
\begin{align}\nonumber
&3\nabla^{\varrho}\nabla_{\varrho}
f_\mathcal{R}-\mathcal{R}\left(\frac{\mathcal{T}}{2}f_\mathcal{Q}-f_\mathcal{R}\right)-\mathcal{T}(8\pi+f_\mathcal{T})+\frac{1}{2}
\nabla^{\varrho}\nabla_{\varrho}(f_\mathcal{Q}\mathcal{T})+\nabla_\varphi\nabla_\varrho(f_\mathcal{Q}\mathcal{T}^{\varphi\varrho})
\\\nonumber
&-2f+(\mathcal{R}f_\mathcal{Q}+4f_\mathcal{T})\mathbb{L}_m+2\mathcal{R}_{\varphi\varrho}\mathcal{T}^{\varphi\varrho}f_\mathcal{Q}
-2g^{\vartheta\xi} \frac{\partial^2\mathbb{L}_m}{\partial
g^{\vartheta\xi}\partial
g^{\varphi\varrho}}\left(f_\mathcal{T}g^{\varphi\varrho}+f_\mathcal{Q}R^{\varphi\varrho}\right)=0.
\end{align}
The strong non-minimal interaction vanishes and reduces to
$f(\mathcal{R},\mathcal{T})$ theory for $f_\mathcal{Q}=0$, while for
the vacuum case (i.e., $f_\mathcal{T}=0$) as well, one can get
$f(\mathcal{R})$ gravity.

Our cosmos is in rapid expansion phase and many stars contain in
non-linear region, yet one can get better understanding of the
formation of massive bodies by analyzing their linear behavior. On
the other hand, this theory produces much complications due to the
factor
$\mathcal{R}_{\varphi\vartheta}\mathcal{T}^{\varphi\vartheta}$, thus
we adopt a separable model of the form \cite{22}
\begin{equation}\label{g5d}
f(\mathcal{R},\mathcal{T},\mathcal{R}_{\varphi\vartheta}\mathcal{T}^{\varphi\vartheta})=\mathcal{R}+\zeta
\mathcal{R}_{\varphi\vartheta}\mathcal{T}^{\varphi\vartheta}.
\end{equation}
There are different acceptable observed values or ranges of the
coupling parameter which produce physically feasible solutions
corresponding to the respective gravity model. Haghani et al.
\cite{22} discussed evolution of the scale factor by studying this
model. The acceptable values of the coupling parameter involving in
this model have also been explored in the framework of isotropic
configuration \cite{22a,22b}.

We consider cylindrically symmetric non-static geometry interior to
the hypersurface as
\begin{equation}\label{g6}
ds^2=-A^2 dt^2+B^2 dr^2+C^2(d\theta^2+\alpha^2 dz^2),
\end{equation}
where $A=A(t,r),~B=B(t,r),~C=C(t,r)$ and $\alpha$ is taken as
constant quantity having dimension of inverse length. However, this
is not the most general cylindrical symmetric line element rather it
is a restricted form. In comoving framework, the four-velocity, heat
flux and four-vector are
\begin{equation}\label{g7}
\mathcal{K}^\varphi=\delta^\varphi_0 A^{-1}, \quad
\varsigma^\varphi=\delta^\varphi_1 \varsigma B^{-1}, \quad
\mathcal{W}^\varphi=\delta^\varphi_1 B^{-1}.
\end{equation}
The factor $\mathcal{Q}$ in this case becomes
\begin{align}\nonumber
\mathcal{Q}&=-\frac{1}{A^3B^3C}\bigg[\mu\big\{2\ddot{C}AB^3+\ddot{B}AB^2C-A''A^2BC-2\dot{A}\dot{C}B^3-2A'C'A^2B\\\nonumber
&+A'B'A^2C-\dot{A}\dot{B}B^2C\big\}+4\varsigma
AB\big\{\dot{C}'AB-A'\dot{C}B-\dot{B}C'A\big\}\\\nonumber
&+P_r\big\{2C''A^3B-\ddot{B}AB^2C+A''A^2BC-2\dot{B}\dot{C}AB^2-2B'C'A^3\\\nonumber
&-A'B'A^2C+\dot{A}\dot{B}B^2C\big\}+\frac{2P_\bot}{C}\big\{C''A^3BC-\ddot{C}AB^3C-\dot{C}^2AB^3\\\nonumber
&-\dot{B}\dot{C}AB^2C+\dot{A}\dot{C}CB^3+C'^2A^3B-B'C'A^3C+A'C'A^2BC\big\}\bigg].
\end{align} Here, dot and prime mean
$\frac{\partial}{\partial t}$ and $\frac{\partial}{\partial r}$,
respectively. The non-vanishing components of field equations
\eqref{g2} for fluid \eqref{g5} corresponding to the model
\eqref{g5d} are given as
\begin{eqnarray}\label{g8}
8\pi\big(\bar{\mu}+\mathcal{T}^{0(\mathcal{C})}_{0})&=&-\frac{1}{B^2}\bigg(\frac{2C''}{C}+\frac{C'^2}{C^2}
-\frac{2C'B'}{CB}\bigg)+\frac{\dot{C}}{CA^2}\bigg(\frac{2\dot{B}}{B}+\frac{\dot{C}}{C}\bigg),\\\label{g8a}
8\pi
(-\bar{\varsigma}+\mathcal{T}^{1(\mathcal{C})}_{0})&=&\frac{1}{AB}\bigg(\frac{2A'\dot
C}{A C}+\frac{2C'\dot B}{C B}-\frac{2\dot C'}{C}\bigg), \\\nonumber
8\pi(\bar{P_r}+\mathcal{T}^{1(\mathcal{C})}_{1})&=&\frac{1}{B^2}\bigg(\frac{C'^2}{C^2}+\frac{2A'C'}{A
C}\bigg)-\frac{1}{A^2}\bigg\{\frac{2\ddot{C}}{C}-\frac{\dot{C}}{C}\bigg(\frac{2\dot{A}}{A}
-\frac{\dot{C}}{C}\bigg)\bigg\},\\\label{g8b}\\\nonumber
8\pi(\bar{P_\bot}+\mathcal{T}^{2(\mathcal{C})}_{2})&=&8\pi{\alpha^2\mathcal{T}^{3(\mathrm{EFF})}_{3}}=
-\frac{1}{A^2}\bigg\{\frac{\dot{B}\dot{C}}{BC} +\frac{\ddot
B}{B}+\frac{\ddot C}{C}-\frac{\dot A}{A}\bigg(\frac{\dot
C}{C}+\frac{\dot B}{B}\bigg)\bigg\}\\\label{g8c}
&+&\frac{1}{B^2}\bigg\{\bigg(\frac{A'}{A}-\frac{B'}{B}\bigg)\frac{C'}{C}
-\frac{A'B'}{A B}+\frac{A''}{A}+\frac{C''}{C}\bigg\},
\end{eqnarray}
where
$\bar{\mu}=\frac{\mu}{1+\zeta\mu},~\bar{\varsigma}=\frac{\varsigma}{1+\zeta\mu},~\bar{P_r}=\frac{P_{r}}{1+\zeta\mu}$
and $\bar{P_\bot}=\frac{P_\bot}{1+\zeta\mu}$. Also, the terms
$\mathcal{T}^{0(\mathcal{C})}_{0},~\mathcal{T}^{1(\mathcal{C})}_{0},~\mathcal{T}^{1(\mathcal{C})}_{1}$
and $\mathcal{T}^{2(\mathcal{C})}_{2}$ are appeared due to the
modified gravity and are given in Appendix $\mathbf{A}$. In this
framework, we deduce the non-vanishing elements of Bianchi identity
with the help of Eq.\eqref{g4a} as
\begin{align}\nonumber
\mathcal{T}^{\varphi\vartheta}_{\quad;\vartheta}\mathcal{K}_{\varphi}&=-\frac{1}{A}\bigg\{\dot{\mu}+2\big(\mu+P_\bot\big)\frac{\dot{C}}{C}
+\big(\mu+P_r\big)\frac{\dot{B}}{B}\bigg\}-\frac{1}{B}\bigg\{\varsigma'+2\varsigma\bigg(\frac{A'}{A}+\frac{C'}{C}\bigg)\bigg\}\\\label{g11}
&=\mathbb{Z}_1-\frac{\zeta}{A(16\pi+\zeta\mathcal{R})}\bigg(\frac{\mu\dot{\mathcal{R}}}{A}+\frac{\varsigma\mathcal{R}'}{B}\bigg),
\end{align}
and
\begin{align}\nonumber
\mathcal{T}^{\varphi\vartheta}_{\quad;\vartheta}\mathcal{W}_{\varphi}&=\frac{1}{A}\bigg\{\dot{\varsigma}+2\varsigma\bigg(\frac{B'}{B}
+\frac{C'}{C}\bigg)\bigg\}+\frac{1}{B}\bigg\{P_r'+2\big(P_r-P_\bot\big)\frac{C'}{C}+\big(\mu+P_r\big)\frac{A'}{A}\bigg\}\\\label{g11a}
&=\mathbb{Z}_2-\frac{\zeta}{B(16\pi+\zeta\mathcal{R})}\bigg(\frac{P_r\mathcal{R}'}{B}+\frac{\varsigma\dot{\mathcal{R}}}{A}\bigg),
\end{align}
where the appearance of $\mathbb{Z}_1$ and $\mathbb{Z}_2$ (provided
in Appendix \textbf{A}) along with modified terms on the right hand
side of the above equations ensure the non-conserved nature of
$f(\mathcal{R},\mathcal{T},\mathcal{Q})$ gravity.

The following equations define the expansion scalar, non-null
components of shear tensor and four-acceleration, respectively as
\begin{align}\label{g11b}
&\Theta=\frac{1}{A}\left(2\frac{\dot C}{C}+\frac{\dot
B}{B}\right),\\\label{g11c} &\sigma_{11}=\frac{2}{3}B^2\sigma, \quad
\sigma_{22}=\frac{\sigma_{33}}{\alpha^2}=-\frac{1}{3}C^2\sigma,\\\label{g11d}
&\sigma^{\varphi\vartheta}\sigma_{\varphi\vartheta}=\frac{2}{3}\sigma^2,~\sigma=\frac{1}{A}\left(\frac{\dot
B}{B}-\frac{\dot C}{C}\right),
\\\label{g11d} &a_{1}=\frac{A'}{A},\quad
a=\sqrt{a^{\varphi}a_{\varphi}}=\frac{A'}{AB}.
\end{align}
We rewrite Eq.\eqref{g8a} to analyze the influence of shear and
expansion scalar on the system as
\begin{equation}\label{g12}
4\pi\left(\bar{\varsigma}-\mathcal{T}^{1(\mathcal{C})}_{0}\right)=\frac{1}{3}\left(\Theta-\sigma\right)'-\sigma
\frac{C'}{C}=\frac{C'}{B}\left[\frac{1}{3}\mathbb{D}_{\mathbb{R}}\left(\Theta-\sigma\right)-\frac{\sigma}{C}\right],
\end{equation}
in which
$\mathbb{D}_{\mathbb{R}}=\frac{1}{C'}\frac{\partial}{\partial r}$
symbolizes the proper radial derivative. The C-energy (which can be
interlinked with the mass function) formula provides the mass of the
cylindrical system as \cite{41ba}
\begin{equation}\label{g13}
\tilde{m}(t,r)=l\hat{\mathrm{E}}=\frac{l}{8}(1-l^{-2}\nabla_\varphi
\hat{r}\nabla^\varphi \hat{r}),
\end{equation}
where $\hat r=\varrho
l,~\varrho^2=\varpi_{(1)\beta}\varpi^{(1)\beta}$ and
$l^2=\varpi_{(2)\beta}\varpi^{(2)\beta}$. Here, $\varrho$ denotes
the circumference radius, $l$ and $\hat {\mathrm{E}}$ are the
specific length and gravitational energy per specific length,
respectively. Moreover, $\varpi_{(1)}=\frac{\partial}{\partial
\theta}$, $\varpi_{(2)}=\frac{\partial}{\partial z}$. Thus the mass
of inner geometry can be calculated as
\begin{equation}\label{g14}
\tilde{m}=\frac{C}{2}\left[\frac{1}{4}-\left(\frac{C'}{B}\right)^2+\left(\frac{\dot
C}{A}\right)^2\right].
\end{equation}

Now, we utilize the suggested form of proper time derivative
($\mathbb{D}_{\mathbb{T}}=\frac{1}{A}\frac{\partial}{\partial t}$)
to discuss the dynamical evolution of corresponding self-gravitating
object. The radius of any compact object decreases continuously
during collapse (as gravity dominates the outward pressure), thus
the velocity of fluid (contained in the interior of that body)
becomes negative, i.e.,
\begin{equation}\label{g15}
\mathbb{U}=\mathbb{D}_{\mathbb{T}}C< 0.
\end{equation}
We figure out the relationship between velocity and C-energy (by
combining Eqs.\eqref{g14} and \eqref{g15}) as
\begin{equation}\label{g16}
\mathbb{E}\equiv\frac{C'}{B}=\left(\mathbb{U}^2+\frac{1}{4}-\frac{2\tilde{m}}{C}\right)^\frac{1}{2}.
\end{equation}
The energy variation inside cylinder can be expressed in terms of
proper time derivative as
\begin{equation}\label{g17}
\mathbb{D}_{\mathbb{T}}\tilde{m}=-4\pi\left[\left(\bar{P_{r}}+\mathcal{T}^{1(\mathcal{C})}_1\right)\mathbb{U}
+\left(\bar{\varsigma}-\mathcal{T}^{1(\mathcal{C})}_0\right)\mathbb{E}\right]C^2+\frac{\dot
C}{8A}.
\end{equation}
In terms of radial derivative, it becomes
\begin{equation}\label{g18}
\mathbb{D}_{\mathbb{R}}\tilde{m}=4\pi\left[\left(\bar{\mu}+\mathcal{T}^{0(\mathcal{C})}_{0}\right)
+\left(\bar{\varsigma}-\mathcal{T}^{1(\mathcal{C})}_{0}\right)\frac{\mathbb{U}}{\mathbb{E}}+\frac{1}{32\pi
C^2}\right]C^2,
\end{equation}
which further leads to
\begin{eqnarray}\nonumber
\frac{3\tilde{m}}{C^3}&=&4\pi\left(\bar{\mu}+\mathcal{T}^{0(\mathcal{C})}_{0}\right)-\frac{4\pi}{C^3}\int^{r}_{0}
C^3\left[\mathbb{D}_{\mathbb{R}}\left(\bar{\mu}+\mathcal{T}^{0(\mathcal{C})}_{0}\right)\right.\\\label{g19}
&-&\left.3\left(\bar{\varsigma}-\mathcal{T}^{1(\mathcal{C})}_{0}\right)\frac{\mathbb{U}}{C\mathbb{E}}\right]C'dr+\frac{3}{8C^2}.
\end{eqnarray}

The last term on the right hand side of the above equations appears
due to the involvement of C-energy \eqref{g13}. The gravitational
field surrounded by a massive body fluctuates which causes the
stretch in nearby celestial structures and can be calculated by the
Weyl tensor $(C^{\lambda}_{\varphi\vartheta\omega})$. This tensor
can completely be defined in terms of two independent components,
namely electric and magnetic parts as
\begin{equation}\nonumber
H_{\varphi\vartheta}=\frac{1}{2}\eta_{\varphi\omega\nu\gamma}C^{\nu\gamma}_{\vartheta\mu}\mathcal{K}^{\omega}
\mathcal{K}^{\mu},\quad
E_{\varphi\vartheta}=C_{\varphi\omega\vartheta\nu}\mathcal{K}^{\omega}\mathcal{K}^{\nu}.
\end{equation}
The alternate expression to calculate the electric part is expressed
as
\begin{equation}\label{g20}
E_{\varphi\vartheta}
=\varepsilon\bigg(\mathcal{W}_{\varphi}\mathcal{W}_{\vartheta}-\frac{h_{\varphi\vartheta}}{3}\bigg),
\end{equation}
where
\begin{eqnarray}\nonumber
\varepsilon &=&\frac{1}{2A^2}\left[\frac{\ddot C}{C}-\frac{\ddot
B}{B}-\left(\frac{\dot C}{C}+\frac{\dot A}{A}\right)\left(\frac{\dot
C}{C}-\frac{\dot
B}{B}\right)\right]+\frac{1}{2B^2}\left[\left(\frac{C'}{C}
-\frac{A'}{A}\right)\right.\\\label{g21}
&\times&\left.\left(\frac{C'}{C}+\frac{B'}{B}\right)+\frac{A''}{A}
-\frac{C''}{C}\right]-\frac{1}{2C^2}.
\end{eqnarray}
The following equation describes how the tidal force affects the
mass of considered geometry as
\begin{equation}\label{g22}
\frac{3\tilde{m}}{C^3}=4\pi\left[\left(\bar{\mu}+\mathcal{T}^{0(\mathcal{C})}_{0}\right)
-\Pi^{(\mathrm{EFF})}\right]+\frac{3}{8C^2}-\varepsilon.
\end{equation}
Here, $\Pi^{(\mathrm{EFF})}=\bar{\Pi}+\Pi^{(\mathcal{C})}$ in which
$\bar{\Pi}=\frac{\Pi}{1+\zeta\mu}$ and
$\Pi^{(\mathcal{C})}=\mathcal{T}^{1(\mathcal{C})}_{1}-\mathcal{T}^{2(\mathcal{C})}_{2}$.

\section{Structure Scalars}

In this section, we split the Riemann curvature tensor orthogonally
(by employing Herrera's approach \cite{41bb}) to calculate modified
structure scalars which are associated with different physical
characteristics of the matter distribution. One can express the
Riemann tensor in terms of the Weyl tensor and trace of the modified
$\mathbb{EMT}$ as
\begin{equation}\label{g23}
\mathcal{R}^{\omega\varphi}_{\gamma\vartheta}=C^{\omega\varphi}_{\gamma\vartheta}+16\pi
\mathcal{T}^{(\mathrm{EFF})[\omega}_{[\gamma}\delta^{\varphi]}_{\vartheta]}+8\pi
\mathcal{T}^{(\mathrm{EFF})}\left(\frac{1}{3}\delta^{\omega}_{[\gamma}\delta^{\varphi}_{\vartheta]}
-\delta^{[\omega}_{[\gamma}\delta^{\varphi]}_{\vartheta]}\right),
\end{equation}
which produces certain tensors like $\mathcal{Y}_{\varphi\vartheta}$
and $\mathcal{X}_{\varphi\vartheta}$ as
\begin{eqnarray}\label{g24}
\mathcal{Y}_{\varphi\vartheta}
&=&\mathcal{R}_{\varphi\omega\vartheta\gamma}\mathcal{K}^{\omega}\mathcal{K}^{\gamma},\\\label{g25}
\mathcal{X}_{\varphi\vartheta}&=&^{\ast}\mathcal{R}^{\ast}_{\varphi\omega\vartheta\gamma}\mathcal{K}^{\omega}\mathcal{K}^{\gamma}
=\frac{1}{2}\eta^{\epsilon\nu}_{\varphi\omega}\mathcal{R}^{\ast}_{\epsilon\nu\vartheta\gamma}\mathcal{K}^{\omega}\mathcal{K}^{\gamma},
\end{eqnarray}
where $\mathcal{R}^{\ast}_{\varphi\omega\vartheta\gamma}
=\frac{1}{2}\eta_{\omega\nu\vartheta\gamma}\mathcal{R}^{\omega\nu}_{\varphi\omega}$
and $\eta^{\epsilon\nu}_{\varphi\omega}$ is the Levi-Civita symbol.
The tensors \eqref{g24} and \eqref{g25} provide four scalars, namely
trace $(\mathcal{Y}_{T},~\mathcal{X}_{T})$ and trace-free components
$(\mathcal{Y}_{TF},~\mathcal{X}_{TF})$ as
\begin{eqnarray}\label{g26}
\mathcal{Y}_{\varphi\vartheta}&=&\frac{h_{\varphi\vartheta}\mathcal{Y}_{T}}{3}+\bigg(\mathcal{W}_{\varphi}\mathcal{W}_{\vartheta}
-\frac{h_{\varphi\vartheta}}{3}\bigg)\mathcal{Y}_{TF},\\\label{g27}
\mathcal{X}_{\varphi\vartheta}&=&\frac{h_{\varphi\vartheta}\mathcal{X}_{T}}{3}+\bigg(\mathcal{W}_{\varphi}\mathcal{W}_{\vartheta}
-\frac{h_{\varphi\vartheta}}{3}\bigg)\mathcal{X}_{TF}.
\end{eqnarray}
The corresponding scalar functions are
\begin{eqnarray}\label{g28}
&&\mathcal{X}_{T}=\frac{8\pi\mu}{1+\zeta\mu}\big(\frac{1}{2}\zeta\mathcal{R}+1\big)+\chi^{(\mathcal{C})}_{1},\\\label{g29}
&&\mathcal{X}_{TF}=-\varepsilon-\frac{4\pi\Pi}{1+\zeta\mu}\big(\frac{1}{2}\zeta\mathcal{R}+1\big),\\\label{g30}
&&\mathcal{Y}_{T}=\frac{1}{1+\zeta\mu}\{4\pi(\mu+3P_{r}-2\Pi)\}\big(\frac{1}{2}\zeta\mathcal{R}+1\big)
+\chi^{(\mathcal{C})}_{2},\\\label{g31}
&&\mathcal{Y}_{TF}=\varepsilon-\frac{4\pi\Pi}{1+\zeta\mu}\big(\frac{1}{2}\zeta\mathcal{R}+1\big)+\chi^{(\mathcal{C})}_{3},
\end{eqnarray}
where $\chi^{(\mathcal{C})}_{1}$,
$\chi^{(\mathcal{\mathcal{C}})}_{2}$ and
$\chi^{(\mathcal{C})}_{3}=\frac{1}{\mathcal{W}_{\varphi}\mathcal{W}_{\vartheta}-\frac{1}{3}h_{\varphi\vartheta}}
\chi_{\varphi\vartheta}^{(\mathcal{C})}$ are provided in Appendix
\textbf{B}. The scalars \eqref{g28} and \eqref{g30} incorporate
energy density of the dynamical cylinder and the local pressure
inside anisotropic distribution along with modified corrections,
respectively. The factor $\mathcal{Y}_{TF}$ can be rewritten by
using Eqs.\eqref{g19} and \eqref{g22} to study structural evolution
as
\begin{align}\nonumber
\mathcal{Y}_{TF}&=-\frac{8\pi\Pi}{1+\zeta\mu}-4\pi\Pi^{(\mathcal{C})}-\frac{2\pi\Pi\zeta\mathcal{R}}{1+\zeta\mu}+\frac{4\pi}{C^3}
\int_0^r
C^3\left[\mathbb{D}_{\mathbb{R}}\left(\bar{\mu}+\mathcal{T}^{0(\mathcal{C})}_{0}\right)\right.\\\label{g32}
&\left.-3\left(\bar{\varsigma}-\mathcal{T}^{1(\mathcal{C})}_{0}\right)\frac{\mathbb{U}}{C
\mathbb{E}}\right]C'dr+\chi^{(\mathcal{C})}_{3}.
\end{align}
This equation confirms the presence of natural variables such as
effective inhomogeneous energy density, heat flux and anisotropic
pressure in modified scalar $\mathcal{Y}_{TF}$. Furthermore, the
term $\mathcal{X}_{TF}$ (appeared in Eq.\eqref{g29}) helps to
measure the inhomogeneity of the dynamical configuration as
\begin{align}\nonumber
\mathcal{X}_{TF}&=4\pi\Pi^{(\mathcal{C})}-\frac{4\pi}{C^3}\int_0^r
C^3\left[\mathbb{D}_{\mathbb{R}}\left(\bar{\mu}+\mathcal{T}^{0(\mathcal{C})}_{0}\right)
-3\left(\bar{\varsigma}-\mathcal{T}^{1(\mathcal{C})}_{0}\right)\frac{\mathbb{U}}{C\mathbb{E}}\right]C'dr\\\label{g33}
&-\frac{2\pi\Pi\zeta\mathcal{R}}{1+\zeta\mu}.
\end{align}

\section{Different Modes of Evolution}

The array of several matter variables (energy density as well as
pressure components) plays an important role in understanding
perplexing nature of our cosmos. It can be noticed from
Eq.\eqref{g32} that $\mathcal{Y}_{TF}$ involves all these quantities
along with heat flux in connection with the modified terms. Thus, in
the case of non-static cylinder, we adopt this modified scalar as
the complexity factor. Henceforth, the system will be
complexity-free if this factor vanishes, i.e., $\mathcal{Y}_{TF}=0$.
To discuss the dynamics of self-gravitating body, we study two
different modes of evolution, namely homologous evolution and
homogeneous expansion. Our goal is to construct some conditions
which will ultimately minimize the complexity throughout the
evolution.

\subsection{Homologous Evolution}

If the system has the same pattern throughout, then this phenomenon
is known as homologous. The collapse of any celestial object occurs
due to inward fall of all the material into its core. On the other
hand, homologous collapse provides direct relation of radial
distance with velocity which means that matter moves towards the
center of an astronomical structure at the same rate during the
collapse. As a result, this kind of collapse emits much more
gravitational radiations in comparison with the body whose core
collapses initially. The alternate form of Eq.\eqref{g12} is given
as
\begin{equation}\label{g34}
\mathbb{D}_{\mathbb{R}}\left(\frac{\mathbb{U}}{C}\right)=\frac{\sigma}{C}+\frac{4\pi}{\mathbb{E}}\left(\bar{\varsigma}
-\mathcal{T}^{1(\mathcal{C})}_{0}\right).
\end{equation}
Its integration provides $\mathbb{U}$ as
\begin{equation}\label{g35}
\mathbb{U}=\mathrm{g}(t)C+C\int^{r}_{0}\left[\frac{\sigma}{C}+\frac{4\pi}{\mathbb{E}}\left(\bar{\varsigma}
-\mathcal{T}^{1(\mathcal{C})}_{0}\right)\right]C'dr.
\end{equation}
Since the above integration is with respect to the radial coordinate
only, thus this equation involves $\mathrm{g}(t)$ as an integration
function. After evaluating this factor at the boundary, we obtain
\begin{equation}\label{g36}
\mathbb{U}=C\left[\frac{\mathbb{U}_{\Sigma}}{C_{\Sigma}}-\int^{r_{\Sigma}}_{r}
\left\{\frac{\sigma}{C}+\frac{4\pi}{\mathbb{E}}\left(\bar{\varsigma}-\mathcal{T}^{1(\mathcal{C})}_{0}\right)\right\}C'dr\right].
\end{equation}
Multiple factors such as heat dissipation and shear scalar play a
significant role in instigating the cylindrical structure to deviate
from homologous mode. We obtain required condition for the system to
be in the state of homologous evolution \cite{42bd} if the terms
inside integral cancel out the effects of each other. Equation
\eqref{g36} is then left with $\mathbb{U}\sim C$ which yields
$\mathbb{U}=\mathrm{g}(t)C$, where
$\mathrm{g}(t)=\frac{\mathbb{U}_{\Sigma}}{C_{\Sigma}}$. In
$f(\mathcal{R},\mathcal{T},\mathcal{Q})$ gravity, the homologous
condition turns out to be
\begin{equation}\label{g37}
\frac{\sigma}{C}+\frac{4\pi
B}{C'}\left(\bar{\varsigma}-\mathcal{T}^{1(\mathcal{C})}_{0}\right)=0.
\end{equation}

\subsection{Homogeneous Expansion}

There is another simplest pattern, known as homogeneous expansion
which requires the condition $\Theta'=0$. Alternate to the former
mode, this phase occurs when the rate at which cosmic entities
expand or collapse is independent of $r$. After combining this
constraint with Eq.\eqref{g12}, we have
\begin{equation}\label{g38}
4\pi\left(\bar{\varsigma}-\mathcal{T}^{1(\mathcal{C})}_{0}\right)=-\frac{C'}{B}
\left[\frac{\sigma}{C}+\frac{1}{3}\mathbb{D}_{\mathbb{R}}(\sigma)\right].
\end{equation}
Using the homologous condition \eqref{g37} in the above equation, it
yields $\mathbb{D}_{\mathbb{R}}(\sigma)=0$. The regularity
conditions in the neighborhood of the center provides $\sigma=0$. As
this solution is obtained by the simultaneous use of Eqs.\eqref{g37}
and \eqref{g38}, thus it must satisfy them. By putting $\sigma=0$ in
Eq.\eqref{g38}, we have the following constraint as
\begin{equation}\label{g39}
\bar{\varsigma}=\mathcal{T}^{1(\mathcal{C})}_{0},
\end{equation}
which demonstrates that the fluid still contains the effects of
dissipation in contrast to $\mathbb{GR}$, as in that case, matter
source was found to be non-dissipative and shear-free corresponding
to homogeneous evolution \cite{33}.

\section{Kinematical and Dynamical Considerations}

Here, we analyze some physical quantities for the simplest possible
mode of evolution. Using homologous condition \eqref{g37} in
Eq.\eqref{g12}, we have
\begin{equation}\label{g40}
\left(\Theta-\sigma\right)'=\left(\frac{3\dot C}{AC}\right)'=0.
\end{equation}
For our convenience, we take $C(t,r)$ as a separable function of $t$
and $r$ which leads to the evolving fluid as geodesic as $A'=0$,
i.e., $a=0$. Therefore, we can take $A=1$ without any loss of
generality. Now we use the values of $\Theta$ and $\sigma$ along
with the geodesic condition ($A=1$) so that we have
\begin{equation}\label{g41}
\Theta-\sigma=\frac{3\dot C}{C},
\end{equation}
which yields $(\Theta-\sigma)'=0$ and hence recover the condition
\eqref{g40}. Thus we can say that the dynamical cylinder will evolve
in homologous mode only if the fluid follows geodesic path and
vice-versa. In $\mathbb{GR}$, no dissipation ($\varsigma=0
\Rightarrow \bar{\varsigma}=0$) means that there is no shear in the
matter source while it does not vanish in this modified gravity as
\begin{equation}\label{g42}
\sigma=4\pi\frac{\mathcal{T}^{1(\mathcal{C})}_{0}BC}{C'}.
\end{equation}
For homogeneous pattern, the shear scalar is obtained by
Eq.\eqref{g38} as
\begin{equation}\label{g43}
\sigma=\frac{\mathrm{h}(t)}{C^3}+\frac{12\pi}{C^3}\int_0^r
BC^3\mathcal{T}^{1(\mathcal{C})}_{0}dr,
\end{equation}
where $\mathrm{h}(t)$ is an arbitrary integration function. We
therefore deduce that homogeneous expansion implies homologous
evolution ($\sigma=0 \Rightarrow \mathbb{U} \sim C$) when
$\bar{\varsigma}=\mathcal{T}^{1(\mathcal{C})}_{0}=0$.

The collapse rate of the current setup (homologous) can be
interlinked with C-energy as
\begin{equation}\label{g44}
\mathbb{D}_{\mathbb{T}}\mathbb{U}=-\frac{\tilde{m}}{C^2}-4\pi
C\left(\bar{P_{r}}+\mathcal{T}^{1(\mathcal{C})}_{1}\right)+\frac{1}{8C},
\end{equation}
which can be rewritten in terms of the scalar $\mathcal{Y}_{TF}$ as
\begin{equation}\label{g45}
\frac{3\mathbb{D}_{\mathbb{T}}\mathbb{U}}{C}=-4\pi\left[\bar{\mu}+\mathcal{T}^{0(\mathcal{C})}_{0}+3\left(\bar{P_{r}}
+\mathcal{T}^{1(\mathcal{C})}_{1}\right)-2\bar{\Pi}-\Pi^{(\mathcal{C})}\right]-\chi_3^{(\mathcal{C})}+\mathcal{Y}_{TF}
+2\pi\zeta\bar{\Pi}\mathcal{R}.
\end{equation}
The modified field equations \eqref{g8}, \eqref{g8b} and \eqref{g8c}
provide
\begin{equation}\label{g46}
4\pi\left[\bar{\mu}+\mathcal{T}^{0(\mathcal{C})}_{0}+3\left(\bar{P_{r}}
+\mathcal{T}^{1(\mathcal{C})}_{1}\right)-2\big(\bar{\Pi}-\Pi^{(\mathcal{C})}\big)\right]=-\frac{2\ddot
C}{C}-\frac{\ddot B}{B},
\end{equation}
and
\begin{equation}\label{g47}
\frac{3\mathbb{D}_{\mathbb{T}}\mathbb{U}}{C}=\frac{3\ddot C}{C}.
\end{equation}
Using Eqs.\eqref{g45}-\eqref{g47}, we obtain
\begin{equation}\label{g48}
\frac{\ddot C}{C}-\frac{\ddot
B}{B}=\mathcal{Y}_{TF}-\chi_3^{(\mathcal{C})}-4\pi\Pi^{(\mathcal{C})}+2\pi\zeta\bar{\Pi}\mathcal{R}.
\end{equation}
The cylindrical configuration becomes free from complexity when
$\frac{\ddot C}{C}-\frac{\ddot
B}{B}+\chi_3^{(\mathcal{C})}+4\pi\Pi^{(\mathcal{C})}-2\pi\zeta\bar{\Pi}\mathcal{R}=0$.

Now, we construct some possible solutions in the presence/absence of
heat dissipation corresponding to complexity-free
($\mathcal{Y}_{TF}=0$) as well as homologous conditions. As we have
taken $A=1$, therefore, we are left with two unknowns, $B(t,r)$ and
$C(t,r)$. For non-dissipative case, these conditions give
\begin{align}\nonumber
&\frac{1}{C^2}+\frac{8\pi}{\zeta}\left(\frac{\zeta\mathcal{R}}{2}+1\right)\bigg[\big(\chi_5\chi_7
+\chi_3\big(\chi_8-1\big)\big)\chi_{12}+\chi_4\big(-\chi_8
\big(\chi_{11}-1\big)+\chi_{11}\\\nonumber
&-\chi_7\chi_{13}-1\big)^{-1}\big\{\chi_5\big(\chi_7
\chi_9+\chi_6\big(\chi_{11}-1\big)\big)+\chi_3\big(\big(\chi_8-1\big)\chi_9-\chi_6\chi_{13}\big)+\chi_1\\\nonumber
&\big(-\chi_8\big(\chi_{11}-1\big)+\chi_{11}-\chi_7\chi_{13}-1\big)\big\}
-\big\{\chi _4 \chi _7 \big(\big(\big(\chi _8-1\big) \big(\chi _4
\big(\chi _{11}-1\big)-\chi _3\\\nonumber &\times\chi
_{12}\big)+\chi _7 \big(\chi _4 \chi _{13}-\chi _5 \chi
_{12}\big)\big) \big(\chi _7 \big(\chi _4 \big(\chi _{14}+\chi
_{15}\big)-\big(\chi _1+\chi _2-1\big) \big(\chi
_{17}-1\big)\big)\\\nonumber &+\big(\chi _6+\chi _7\big) \big(\chi
_4 \chi _{16}-\chi _3 \big(\chi _{17}-1\big)\big)\big)-\big(\chi _7
\big(\chi _4 \big(\chi _9+\chi _{10}\big)-\big(\chi _1+\chi
_2-1\big) \chi _{12}\big)\\\nonumber &+\big(\chi _6+\chi _7\big)
\big(\chi _4 \big(\chi _{11}-1\big)-\chi _3 \chi _{12}\big)\big)
\big(\big(\chi _8-1\big) \big(\chi _4 \chi _{16}-\chi _3 \big(\chi
_{17}-1\big)\big)+\chi _7 \big(\chi _4 \\\nonumber &\times\chi
_{18}-\chi _5 \big(\chi
_{17}-1\big)\big)\big)\big)\big\}^{-1}\big\{\big(\chi _3 \chi
_6+\chi _1 \chi _7+\chi _2 \chi _7+\chi _3 \chi _7-\chi
_7-\big(\big(\chi _5 \chi _7+\chi _3 \\\nonumber &\times\big(\chi
_8-1\big)\big) \chi _{12}+\chi _4 \big(-\chi _8 \big(\chi
_{11}-1\big)+\chi _{11}-\chi _7 \chi
_{13}-1\big)\big)^{-1}\big(\big(\chi _5 \chi _7+ \big(\chi
_8-1\big)\\\nonumber &\times\chi _3\big) \big(\big(\big(\chi _1+\chi
_2-1\big) \chi _7+\chi _3 \big(\chi _6+\chi _7\big)\big) \chi
_{12}-\chi _4 \big(\chi _6 \big(\chi _{11}-1\big)+\chi _7 \big(\chi
_9+\chi _{10}\\\nonumber &+\chi
_{11}-1\big)\big)\big)\big)\big)\big(\big(\big(-\chi _5 \chi _7-\chi
_3 \big(\chi _8-1\big)\big) \chi _{12}+\chi _4 \big(\chi _8
\big(\chi _{11}-1\big)-\chi _{11}+\chi _7 \chi _{13}\\\nonumber
&+1\big)\big) \big(\chi _4 \big(\chi _7 \chi _{14}+\chi _6 \chi
_{16}\big)-\chi _3 \chi _6 \big(\chi _{17}-1\big)-\chi _1 \chi _7
\big(\chi _{17}-1\big)\big)-\big(\chi _4 \big(\chi _7 \chi
_9\\\nonumber &+\chi _6 \big(\chi _{11}-1\big)\big)-\big(\chi _3
\chi _6+\chi _1 \chi _7\big) \chi _{12}\big) \big(-\chi _5 \chi _7
\big(\chi _{17}-1\big)- \big(\chi _8-1\big) \big(\chi
_{17}-1\big)\\\nonumber &\times\chi _3+\chi _4 \big(\big(\chi
_8-1\big) \chi _{16}+\chi _7 \chi
_{18}\big)\big)\big)\big\}+\big\{\big(-\chi _5 \chi _7-\chi _3
\big(\chi _8-1\big)\big) \chi _{12}+\chi _4 \big(\chi _8\\\nonumber
&\times\big(\chi _{11}-1\big)-\chi _{11}+\chi _7 \chi
_{13}+1\big)\big\}^{-1}\big\{\big(\chi _1 \big(\chi _8-1\big)-\chi
_5 \chi _6\big) \chi _{12}+\chi _4 \big(\big(1-\chi _8\big)
\\\nonumber &\times\chi _9+\chi _6 \chi _{13}\big)+\big(\big(\big(\chi _8-1\big) \big(\chi _4
\big(\chi _{11}-1\big)-\chi _3 \chi _{12}\big)+\chi _7\big(\chi _4
\chi _{13}-\chi _5 \chi _{12}\big)\big) \big(\chi _7\\\nonumber
&\times \big(\chi _4 \big(\chi _{14}+\chi _{15}\big)-\big(\chi
_1+\chi _2-1\big) \big(\chi _{17}-1\big)\big)+\big(\chi _6+\chi
_7\big) \big(\chi _4 \chi _{16}- \big(\chi _{17}-1\big)\\\nonumber
&\times\chi _3\big)\big)-\big(\chi _7 \big(\chi _4 \big(\chi _9+\chi
_{10}\big)-\big(\chi _1+\chi _2-1\big) \chi _{12}\big)+\big(\chi
_6+\chi _7\big) \big(\chi _4 \big(\chi _{11}-1\big)\\\nonumber
&-\chi _3 \chi _{12}\big)\big) \big(\big(\chi _8-1\big) \big(\chi _4
\chi _{16}-\chi _3 \big(\chi _{17}-1\big)\big)+\chi _7 \big(\chi _4
\chi_{18}-\chi_5\big(\chi_{17}-1\big)\big)\big)\big)^{-1}\\\nonumber
&\times\big(\big(\big(\chi _5 \chi _6+\chi _5 \chi _7-\chi _1
\big(\chi _8-1\big)-\chi _2 \big(\chi _8-1\big)+\chi _8-1\big) \chi
_{12}+\chi _4 \big(\big(\chi _8-1\big) \chi _9\\\nonumber
&+\big(\chi _8-1\big) \chi _{10}-\big(\chi _6+\chi _7\big) \chi
_{13}\big)\big) \big(\big(\big(-\chi _5 \chi _7-\chi _3 \big(\chi
_8-1\big)\big) \chi _{12}+\chi _4 \big( \big(\chi
_{11}-1\big)\\\nonumber &\times\chi _8-\chi _{11}+\chi _7 \chi
_{13}+1\big)\big) \big(\chi _4 \big(\chi _7 \chi _{14}+\chi _6 \chi
_{16}\big)-\chi _3 \chi _6 \big(\chi _{17}-1\big)- \big(\chi
_{17}-1\big)\\\nonumber &\times\chi _1 \chi _7\big)-\big(\chi _4
\big(\chi _7 \chi _9+\chi _6 \big(\chi _{11}-1\big)\big)-\big(\chi
_3 \chi _6+\chi _1 \chi _7\big) \chi _{12}\big) \big(-\chi _5 \chi
_7 \big(\chi _{17}-1\big)\\\nonumber &-\chi _3 \big(\chi _8-1\big)
\big(\chi _{17}-1\big)+\chi _4 \big(\big(\chi _8-1\big) \chi
_{16}+\chi _7 \chi _{18}\big)\big)\big)\big)
\big\}\bigg]-\bigg\{\frac{\dot{C}\big(\dot{B}C-B\dot{C}\big)}{BC^2}\\\label{g51}
&+\frac{C\big(B'C'-BC''\big)+BC'^2}{B^3
C^2}-\frac{\ddot{B}}{B}+\frac{\ddot{C}}{C}\bigg\}=0,\\\nonumber
&-\chi _7 \big(\chi _4 \chi _9-\chi _1 \chi _{12}\big)-\chi _6
\big(\chi _4 \big(\chi _{11}-1\big)-\chi _3 \chi
_{12}\big)+\big\{\big(\big(\chi _8-1\big) \big(\chi _4 \big(\chi
_{11}-1\big)\\\nonumber &-\chi _3 \chi _{12}\big)+\chi _7 \big(\chi
_4 \chi _{13}-\chi _5 \chi _{12}\big)\big) \big(\chi _7 \big(\chi _4
\big(\chi _{14}+\chi _{15}\big)-\big(\chi _1+\chi _2-1\big)
\big(\chi _{17}-1\big)\big)\\\nonumber &+\big(\chi _6+\chi _7\big)
\big(\chi _4 \chi _{16}-\chi _3 \big(\chi
_{17}-1\big)\big)\big)-\big(\chi _7 \big(\chi _4 \big(\chi _9+\chi
_{10}\big)-\big(\chi _1+\chi _2-1\big) \chi _{12}\big)\\\nonumber
&+\big(\chi _6+\chi _7\big) \big(\chi _4 \big(\chi _{11}-1\big)-\chi
_3 \chi _{12}\big)\big) \big(\big(\chi _8-1\big) \big(\chi _4 \chi
_{16}-\chi _3 \big(\chi _{17}-1\big)\big)+\chi _7 \big(\chi
_4\\\nonumber &\times \chi _{18}-\chi _5 \big(\chi
_{17}-1\big)\big)\big)\big\}^{-1}\big\{\big(\chi _7 \big(\chi _4
\big(\chi _9+\chi _{10}\big)-\big(\chi _1+\chi _2-1\big) \chi
_{12}\big)+\big(\chi _6\\\nonumber &+\chi _7\big) \big(\chi _4
\big(\chi _{11}-1\big)-\chi _3 \chi _{12}\big)\big)
\big(\big(\big(-\chi _5 \chi _7-\chi _3 \big(\chi _8-1\big)\big)
\chi _{12}+\chi _4 \big(\chi _8 \big(\chi _{11}-1\big)\\\nonumber
&-\chi _{11}+\chi _7 \chi _{13}+1\big)\big) \big(\chi _4 \big(\chi
_7 \chi _{14}+\chi _6 \chi _{16}\big)-\chi _3 \chi _6 \big(\chi
_{17}-1\big)-\chi _1 \chi _7 \big(\chi _{17}-1\big)\big)\\\nonumber
&-\big(\chi _4 \big(\chi _7 \chi _9+\chi _6 \big(\chi
_{11}-1\big)\big)-\big(\chi _3 \chi _6+\chi _1 \chi _7\big) \chi
_{12}\big) \big(-\chi _5 \chi _7 \big(\chi _{17}-1\big)-\chi _3
\big(\chi _8\\\label{g52} &-1\big) \big(\chi _{17}-1\big)+\chi _4
\big(\big(\chi _8-1\big) \chi _{16}+\chi _7 \chi
_{18}\big)\big)\big)\big\}=0.
\end{align}
In the presence of dissipation, the complexity-free condition
remains the same as given in Eq.\eqref{g51} while the homologous
condition takes the form
\begin{align}\nonumber
\bar{\varsigma}&=\frac{\chi _6}{1-\chi
_8}\bigg[\frac{1}{\zeta}+\big\{\big(\big(\chi _8-1\big) \big(\chi _4
\big(\chi _{11}-1\big)-\chi _3 \chi _{12}\big)+\chi _7\big(\chi _4
\chi _{13}-\chi _5 \chi _{12}\big)\big) \big(\chi _7
\\\nonumber &\times \big(\chi _4 \big(\chi _{14}+\chi _{15}\big)-\big(\chi _1+\chi
_2-1\big) \big(\chi _{17}-1\big)\big)+\big(\chi _6+\chi _7\big)
\big(\chi _4 \chi _{16}-\chi _3 \big(\chi _{17}\\\nonumber
&-1\big)\big)\big)-\big(\chi _7 \big(\chi _4 \big(\chi _9+\chi
_{10}\big)-\big(\chi _1+\chi _2-1\big) \chi _{12}\big)+\big(\chi
_6+\chi _7\big) \big(\chi _4 \big(\chi _{11}-1\big)\\\nonumber
&-\chi _3 \chi _{12}\big)\big) \big(\big(\chi _8-1\big) \big(\chi _4
\chi _{16}-\chi _3 \big(\chi _{17}-1\big)\big)+\chi _7 \big(\chi _4
\chi _{18}-\chi _5 \big(\chi
_{17}-1\big)\big)\big)\big\}^{-1}\\\nonumber
&\times\big\{\big(\big(\chi _5 \chi _7+\chi _3 \big(\chi
_8-1\big)\big) \chi _{12}+\chi _4 \big(-\chi _8 \big(\chi
_{11}-1\big)+\chi _{11}-\chi _7 \chi _{13}-1\big)\big) \big(\chi
_4\\\nonumber &\times \big(\chi _7 \chi _{14}+\chi _6 \chi
_{16}\big)-\chi _3 \chi _6 \big(\chi _{17}-1\big)-\chi _1 \chi _7
\big(\chi _{17}-1\big)\big)+\big(\chi _4 \big(\chi _7 \chi _9+\chi
_6 \big(\chi _{11}\\\nonumber &-1\big)\big)-\big(\chi _3 \chi
_6+\chi _1 \chi _7\big) \chi _{12}\big) \big(-\chi _5 \chi _7
\big(\chi _{17}-1\big)-\chi _3 \big(\chi _8-1\big) \big(\chi
_{17}-1\big)+\chi _4\\\label{g53} &\times \big(\big(\chi _8-1\big)
\chi_{16}+\chi_7\chi_{18}\big)\big)\big\}\bigg]+\frac{C'\big(B\dot{C}-\dot{B}C\big)}{4\pi
B^2C^2}.
\end{align}
The values of $\chi_i$, $i=1,2,3,...,18$ are provided in Appendix
\textbf{B}.

\section{Stability of Zero Complexity Condition}

It is possible that a system that initiates with complexity-free
interior develops complex nature at a later time, i.e., the
condition of $\mathcal{Y}_{TF}=0$ may be disturbed during the
evolution of the system. In this section, we investigate factors
which may affect the stability. For this, we derive the evolution
equation corresponding to the cylindrical geometry by using
Eqs.\eqref{g11}, \eqref{g29} and \eqref{g31} through the procedure
\cite{41bb}
\begin{align}\nonumber
&-4\pi\left(\bar{\mu}+\bar{p_{r}}\right)\sigma-\frac{4\pi}{B}\bigg(\bar{\varsigma}'-\frac{\bar{\varsigma}
C'}{C}-\frac{3C'\mathcal{T}^{1(\mathcal{C})}_{0}}{C}\bigg)-\dot{\mathcal{Y}}_{TF}+\dot\chi_3^{(\mathcal{C})}
-8\pi\dot{\bar{\Pi}}-4\pi\mathbb{Z}_1\\\nonumber&+4\pi\dot{\mathcal{T}}^{0(\mathcal{C})}_{0}-2\pi\zeta\big(\bar{\Pi}\mathcal{R}\big)^.
+\frac{\zeta}{16\pi+\zeta\mathcal{R}}\big(\mu\dot{\mathcal{R}}+\frac{\bar{\varsigma}\mathcal{R}'}{B}\big)
-\frac{3\dot{C}}{C}\big(\mathcal{Y}_{TF}+2\pi\zeta\bar{\Pi}\mathcal{R}\\\label{g54}
&-\chi_3^{(\mathcal{C})}\big)+\frac{12\pi\dot{C}}{C}\left(\mathcal{T}^{0(\mathcal{C})}_{0}+\mathcal{T}^{1(\mathcal{C})}_{1}
-\bar{\Pi}^{(\mathcal{C})}\right)-\frac{16\pi\bar{\Pi}\dot{C}}{C}=0.
\end{align}
Firstly, we assume the non-dissipative case and take
$\mathcal{Y}_{TF}=0=\Pi^{(\mathrm{EFF})}=\varsigma=\sigma$ at
initial time (say $t=0$) in the above equation so that we have
\begin{align}\nonumber
&\frac{12\pi{C'}\mathcal{T}^{1(\mathcal{C})}_{0}}{BC}-\dot{\mathcal{Y}}_{TF}+\dot\chi_3^{(\mathcal{C})}
+\frac{\zeta\mu\dot{\mathcal{R}}}{16\pi+\zeta\mathcal{R}}
+\frac{3\dot{C}\chi_3^{(\mathcal{C})}}{C}
-8\pi\dot{\bar{\Pi}}-4\pi\mathbb{Z}_1+4\pi\dot{\mathcal{T}}^{0(\mathcal{C})}_{0}\\\label{g55}
&-2\pi\zeta\big(\bar{\Pi}\mathcal{R}\big)^.+\frac{12\pi\dot{C}}{C}\left(\mathcal{T}^{0(\mathcal{C})}_{0}
+\mathcal{T}^{1(\mathcal{C})}_{1}\right)=0.
\end{align}
Using this equation in the derivative of Eq.\eqref{g32} (at $t=0$)
to eliminate $\dot{\mathcal{Y}}_{TF}$, we obtain
\begin{align}\nonumber
&4\pi\frac{\partial}{\partial
t}\bigg[\frac{1}{C^3}\int^{r}_{0}C^3\bigg\{(\bar{\mu}+\mathcal{T}^{0(\mathcal{C})}_{0})'
+\frac{3\mathcal{T}^{1(\mathcal{C})}_{0}B\dot{C}}{C}\bigg\}dr\bigg]=4\pi\dot{\mathcal{T}}^{0(\mathcal{C})}_{0}-4\pi\mathbb{Z}_1\\\nonumber
&+2\pi\zeta\big(\bar{\Pi}\mathcal{R}\big)^.+4\pi\dot{\Pi}^{(\mathcal{C})}+\frac{12\pi{C'}\mathcal{T}^{1(\mathcal{C})}_{0}}{BC}
+\frac{\zeta\mu\dot{\mathcal{R}}}{16\pi+\zeta\mathcal{R}}+\frac{3\dot{C}\chi_3^{(\mathcal{C})}}{C}\\\label{g56}
&+\frac{12\pi\dot{C}}{C}\left(\mathcal{T}^{0(\mathcal{C})}_{0}
+\mathcal{T}^{1(\mathcal{C})}_{1}\right).
\end{align}
It is known that the stability of vanishing complexity depends on
state variables such as pressure and energy density. Thus,
inhomogeneity and anisotropy in energy density and pressure,
respectively trigger complexity in the system.

In the most general case, when the system is dissipative,
Eq.\eqref{g54} with $\mathcal{Y}_{TF}=0=\Pi^{(\mathrm{EFF})}=\sigma$
yields
\begin{align}\nonumber
&-\frac{4\pi}{B}\bigg(\bar{\varsigma}'-\frac{\bar{\varsigma}
C'}{C}-\frac{3C'\mathcal{T}^{1(\mathcal{C})}_{0}}{C}\bigg)-\dot{\mathcal{Y}}_{TF}+\dot\chi_3^{(\mathcal{C})}
-8\pi\dot{\bar{\Pi}}-4\pi\mathbb{Z}_1\\\nonumber&+4\pi\dot{\mathcal{T}}^{0(\mathcal{C})}_{0}-2\pi\zeta\big(\bar{\Pi}\mathcal{R}\big)^.
+\frac{\zeta}{16\pi+\zeta\mathcal{R}}\bigg(\mu\dot{\mathcal{R}}+\frac{\bar{\varsigma}\mathcal{R}'}{B}\bigg)
+\frac{3\dot{C}\chi_3^{(\mathcal{C})}}{C}\\\label{g57}
&+\frac{12\pi\dot{C}}{C}\left(\mathcal{T}^{0(\mathcal{C})}_{0}+\mathcal{T}^{1(\mathcal{C})}_{1}
\right)=0.
\end{align}
It can clearly be deduced from the above equation that now the heat
flux also affects the stability of the $\mathcal{Y}_{TF}=0$
condition.

\section{Final Remarks}

The complex self-gravitating stellar bodies have prompted many
researchers to study the cosmos due to their interesting behavior.
In this paper, we have analyzed the effects of matter-geometry
interaction on non-static cylindrical geometry whose interior
involves inhomogeneous energy density, dissipative flux and
anisotropic pressure. We have used Herrera's technique to split the
Riemann curvature tensor and obtained four scalars, each of them is
linked with some particular physical properties. We have chosen
$\mathcal{Y}_{TF}$ as the best candidate for the complexity factor
due to following reasons.
\begin{enumerate}
\item It has been treated as the complexity factor in $\mathbb{GR}$ \cite{31} as well
as
$f(\mathcal{R},\mathcal{T},\mathcal{R}_{\varphi\vartheta}\mathcal{T}^{\varphi\vartheta})$
gravity \cite{26c} for static spacetime and hence can be retrieved
from Eq.\eqref{g31} for the static scenario.
\item Only this factor incorporates all the quantities, i.e, energy density
inhomogeneity, pressure anisotropy and heat dissipation together
with modified corrections that can cause to make the system more
complex.
\end{enumerate}
We have studied the structural evolution of dynamical cylinder with
the help of two different evolutionary patterns, namely homogeneous
expansion and homologous evolution. Further, we have adopted
$\mathcal{Y}_{TF}=0$ together with homologous condition \eqref{g37}
to find the corresponding solution in the presence and absence of
dissipative flux. We have discussed some particular factors which
deviate the system from complexity-free scenario throughout the
evolution.

The presence of correction terms due to the model \eqref{g5d}
produce more complexity in the dynamical cylinder, as the matter
variables along with metric potentials are involved in modified
field equations. We have considered the geodesic nature of
homologous fluid, i.e., $A=1$, which leads the homologous pattern to
be the simplest mode of dynamical evolution. In $\mathbb{GR}$, the
condition $\mu=\varsigma=\Pi=0$ leads to $\mathcal{Y}_{TF}=0$, while
an additional condition
($8\pi\bar{\Pi}+4\pi\Pi^{(\mathcal{C})}+2\pi\zeta\mathcal{R}\bar{\Pi}-\chi_3^{(\mathcal{C})}=0$)
is required to obtain the complexity-free structure in this gravity.
Here, the non-dissipative case ($\varsigma=0$) does not provide
shear-free system which is inconsistent with $\mathbb{GR}$. This
phenomenon has been discussed for minimal as well as non-minimal
$f(\mathcal{R},\mathcal{T})$ models from which it is observed that
simplest modes of evolution are found to be compatible with each
other for the case of minimal coupling, and otherwise, not
\cite{36b}. For the considered
$f(\mathcal{R},\mathcal{T},\mathcal{Q})$ model, our obtained results
are compatible with those of $f(\mathcal{R},\mathcal{T})$ theory.
Finally, we have discussed the stability condition of the
complexity-free factor for dissipative as well as non-dissipative
models. The factors due to which the system can depart from zero
complexity throughout the evolution have also been addressed. It is
found that the dynamical cylinder becomes more complex in the
framework of modified model \eqref{g5d} as compared to
$\mathbb{GR}$. All our findings reduce to \cite{33} by choosing
$\zeta=0$.

\vspace{0.25cm}

\section*{Appendix A}

\renewcommand{\theequation}{A\arabic{equation}}
\setcounter{equation}{0} The correction terms in the field equations
\eqref{g8}-\eqref{g8c} are
\begin{align}\nonumber
\mathcal{T}^{(\mathcal{C})}_{00}&=\frac{\zeta}{8\pi(1+\zeta\mu)}\bigg[\mu\bigg(\frac{3\ddot{B}}{2B^2}-\frac{3\dot{A}\dot{B}}{2AB}
+\frac{3\ddot{C}}{C}-\frac{\dot{A}\dot{C}}{AC}-\frac{\dot{C}^2}{C^2}-\frac{3AA''}{2B^2}+\frac{2A'^2}{B^2}\\\nonumber
&-\frac{1}{2}A^2\mathcal{R}-\frac{AA'B'}{2B^3}-\frac{2\dot{B}\dot{C}}{BC}-\frac{3AA'C'}{B^2C}\bigg)-\dot{\mu}\bigg(\frac{3\dot{A}}{A}
+\frac{\dot{C}}{C}+\frac{\dot{B}}{2B}\bigg)+\frac{\mu''A^2}{2B^2}\\\nonumber
&-\mu'\bigg(\frac{A^2B'}{2B^3}-\frac{A^2C'}{B^2}\bigg)+P_{r}\bigg(-\frac{\ddot{B}}{2B}+\frac{AA''}{2B^2}-\frac{AA'B'}{2B^3}-\frac{A^2C''}{B^2C}
-\frac{A^2C'^2}{B^2C^2}\\\nonumber
&+\frac{2A^2B'C'}{B^3C}-\frac{\dot{B}\dot{C}}{BC}-\frac{2A^2B'\dot{C}}{B^3C}+\frac{2A'^2}{B^2}\bigg)+\frac{\dot{P}_{r}\dot{B}}{2B}
+P'_{r}\bigg(\frac{A^2B'}{2B^3}-\frac{2A^2C'}{B^2C}\bigg)\\\nonumber
&-\frac{P''_{r}A}{2B^2}-P_{\bot}\bigg(\frac{\ddot{C}}{C}+\frac{\dot{C}^2}{C}-\frac{\dot{A}\dot{C}}{AC}
-\frac{AA'C'}{B^2C}-\frac{A^2C''}{B^2C}+\frac{A^2B'C'}{B^3C}-\frac{A^2C'^2}{B^2C^2}\\\nonumber
&+\frac{\dot{B}\dot{C}}{BC}\bigg)-\frac{3\dot{P}_{\bot}\dot{C}}{C}+\frac{P'_{\bot}A^2C'}{B^2C}-\varsigma\bigg(\frac{9\dot{A}A'}{2AB}
-\frac{2A\dot{C}'}{BC}+\frac{3A'\dot{C}}{BC}+\frac{5A\dot{B}C'}{B^2C}\\\label{A1}
&+\frac{3A'\dot{B}}{B^2}+\frac{\dot{A}C'}{BC}+\frac{2A\dot{C}C'}{BC^2}\bigg)+\frac{2\dot{\varsigma}A'}{B}-\frac{2\varsigma'A\dot{C}}{BC}
+\frac{A^2\mathcal{Q}}{2}\bigg],\\\nonumber
\mathcal{T}^{(\mathcal{C})}_{01}&=\frac{\zeta}{8\pi(1+\zeta\mu)}\bigg[\mu\bigg(-\frac{\dot{A}A'}{A^2}-\frac{3\dot{A}'}{2A}
+\frac{2\dot{C}'}{C}-\frac{2\dot{B}C'}{BC}-\frac{2A'\dot{C}}{AC}\bigg)-\frac{3\dot{\mu}A'}{2A}\\\nonumber
&+\mu'\bigg(\frac{\dot{C}}{C}-\frac{\dot{B}}{2B}\bigg)+\frac{\dot{\mu}'}{2}+P_{r}\bigg(\frac{A'^2}{2B^2}-\frac{A'\dot{B}}{2AB}
-\frac{2\dot{C}'}{C}+\frac{2\dot{B}C'}{BC}+\frac{2A'\dot{C}}{AC}\bigg)\\\nonumber
&+\dot{P}_{r}\bigg(\frac{A'}{2A}-\frac{C'}{C}\bigg)+\frac{P'_{r}\dot{B}}{2B}-\frac{\dot{P}'_{r}}{2}+\frac{\dot{P}_{\bot}C'}{C}
+\frac{P'_{\bot}\dot{C}}{C}+\varsigma\bigg(\frac{1}{2}AB\mathcal{R}-\frac{2B\ddot{C}}{AC}\\\label{A2}
&-\frac{\ddot{B}}{A}+\frac{A''}{B}+\frac{2B\dot{A}\dot{C}}{A^2C}+\frac{2AC''}{BC}-\frac{2AB'C'}{B^2C}+\frac{\dot{A}\dot{B}}{A^2}
-\frac{A'B'}{B^2}\bigg)\bigg],\\\nonumber
\mathcal{T}^{(\mathcal{C})}_{11}&=\frac{\zeta}{8\pi(1+\zeta\mu)}\bigg[\mu\bigg(\frac{B\ddot{B}}{2A^2}-\frac{B\dot{A}\dot{B}}{2A^3}
-\frac{B^2\ddot{C}}{A^2C}+\frac{B^2\dot{A}\dot{C}}{A^3C}-\frac{B^2\dot{C}^2}{A^2C^2}-\frac{A''}{2A}+\frac{A'B'}{2AB}\\\nonumber
&-\frac{A'C'}{AC}\bigg)+\dot{\mu}\bigg(\frac{B^2\dot{A}}{2A^3}-\frac{2B^2\dot{C}}{A^2C}\bigg)
+\frac{\mu'A'}{2A}-\frac{\ddot{\mu}B^2}{2A^2}+P_{r}\bigg(\frac{1}{2}B^2\mathcal{R}+\frac{3B\dot{A}\dot{B}}{2A^3}\\\nonumber
&-\frac{3B\ddot{B}}{2A^2}-\frac{2A'C'}{AC}+\frac{3A''}{2A}-\frac{3A'B'}{2AB}-\frac{2B'C'}{BC}
+\frac{3C''}{C}-\frac{3B\dot{B}\dot{C}}{A^2C}-\frac{2B'\dot{C}}{BC}\\\nonumber
&-\frac{C'^2}{C^2}\bigg)+\dot{P}_{r}\bigg(\frac{B^2\dot{C}}{A^2C}-\frac{B^2\dot{A}}{2A^3}\bigg)
-P'_{r}\bigg(\frac{A'}{2A}+\frac{C'}{C}\bigg)+\frac{\ddot{P}_{r}B^2}{2A^2}+P_{\bot}\bigg(\frac{A'C'}{AC}\\\nonumber
&-\frac{B^2\ddot{C}}{A^2C}-\frac{B^2\dot{C}^2}{A^2C^2}+\frac{B^2\dot{A}\dot{C}}{A^3C}+\frac{C''}{C}
-\frac{B'C'}{BC}+\frac{C'^2}{C^2}-\frac{B\dot{B}\dot{C}}{A^2C}\bigg)-\frac{\dot{P}_{\bot}B^2\dot{C}}{A^2C}\\\nonumber
&-\frac{P'_{\bot}C'}{C}+\varsigma\bigg(\frac{2B\dot{C}'}{AC}-\frac{B\dot{A}C'}{A^2C}
-\frac{3BA'\dot{C}}{A^2C}-\frac{2B\dot{C}C'}{AC^2}-\frac{4\dot{B}C'}{AC}\bigg)+\frac{B^2\mathcal{Q}}{2}\\\label{A3}
&-\frac{2\dot{\varsigma}BC'}{AC}\bigg],\\\nonumber
\mathcal{T}^{(\mathcal{C})}_{22}&=\frac{\zeta}{8\pi(1+\zeta\mu)}\bigg[\mu\bigg(-\frac{C^2\ddot{B}}{2A^2B}+\frac{C^2\dot{A}\dot{B}}{2A^3B}
-\frac{C^2A''}{2AB^2}+\frac{C^2A'B'}{2AB^3}-\frac{C\dot{B}\dot{C}}{A^2B}\\\nonumber
&-\frac{CA'C'}{AB^2}\bigg)+\dot{\mu}\bigg(\frac{C^2\dot{A}}{2A^3}-\frac{C^2\dot{B}}{A^2B}-\frac{C\dot{C}}{A^2}\bigg)
-\frac{\mu'C^2A'}{2AB^2}-\frac{\ddot{\mu}C^2}{2A^2}+P_{r}\bigg(\frac{C^2\dot{A}\dot{B}}{2A^3B}\\\nonumber
&-\frac{C^2\ddot{B}}{2A^2B}-\frac{CA'C'}{AB^2}-\frac{C^2A''}{2AB^2}+\frac{C^2A'B'}{2AB^3}+\frac{CB'C'}{B^3}-\frac{C\dot{B}\dot{C}}{A^2B}
-\frac{2CB'\dot{C}}{B^3}\bigg)\\\nonumber
&-\frac{\dot{P}_{r}C^2\dot{B}}{2A^2B}+P'_{r}\bigg(-\frac{C^2A'}{AB^2}+\frac{C^2B'}{2B^3}-\frac{CC'}{B^2}\bigg)-\frac{P''_{r}C^2}{2B^2}
+P_{\bot}\bigg(\frac{1}{2}C^2\mathcal{R}\\\nonumber
&+\frac{2CC''}{B^2}-\frac{2C\ddot{C}}{A^2}-\frac{2\dot{C}^2}{A^2}+\frac{2C'^2}{B^2}-\frac{2CB'C'}{B^3}+\frac{2CA'C'}{AB^2}
-\frac{2C\dot{B}\dot{C}}{A^2B}-2\\\nonumber
&+\frac{2C\dot{A}\dot{C}}{A^3}\bigg)+\dot{P}_{\bot}C^2\bigg(\frac{\dot{B}}{2A^2B}-\frac{\dot{A}}{2A^3}\bigg)+P'_{\bot}C^2\bigg(-\frac{A'}{2AB^2}
+\frac{B'}{2B^3}\bigg)+\frac{\ddot{P}_{\bot}C^2}{2A^2}\\\nonumber
&-\frac{P''_{\bot}C^2}{2B^2}+\varsigma\bigg(-\frac{C^2\dot{A}'}{A^2B}+\frac{C^2\dot{A}A'}{A^3B}-\frac{C\dot{A}C'}{A^2B}+\frac{C^2\dot{B}B'}{AB^3}
-\frac{C^2\dot{B}'}{AB^2}-\frac{CA'\dot{C}}{A^2B}\\\label{A4}
&-\frac{2C\dot{B}C'}{AB^2}\bigg)-\dot{\varsigma}\bigg(\frac{C^2A'}{A^2B}+\frac{CC'}{AB}\bigg)-\varsigma'\bigg(\frac{C^2\dot{B}}{AB^2}
+\frac{C\dot{C}}{AB}\bigg)-\frac{\dot{\varsigma}'C^2}{AB}+\frac{C^2\mathcal{Q}}{2}\bigg].
\end{align}
The terms $\mathbb{Z}_1$ and $\mathbb{Z}_2$ in Eqs.\eqref{g11} and
\eqref{g11a} are
\begin{align}\nonumber
\mathbb{Z}_1&=\frac{2\zeta}{16\pi+\zeta\mathcal{R}}\bigg[\big(\frac{\varsigma
B\mathcal{R}^{10}}{A}\big)^.+\big(\frac{\varsigma
B\mathcal{R}^{11}}{A}\big)'-\big(\mu\mathcal{R}^{00}\big)^.-\big(\mu\mathcal{R}^{01}\big)'+\dot{\mu}\mathcal{G}^{00}\\\nonumber
&+\mu'\mathcal{G}^{01}+\frac{1}{2A^2}\bigg\{\mathcal{R}_{00}\bigg(\frac{\dot{\mu}}{A^2}-\frac{2\mu\dot{A}}{A^3}\bigg)+2\mathcal{R}_{01}
\bigg(\frac{\dot{\varsigma}}{AB}-\frac{\varsigma\dot{A}}{A}-\frac{\varsigma\dot{B}}{B}\bigg)\\\label{A5}
&+\mathcal{R}_{11}\bigg(\frac{\dot{P}_r}{B^2}
-\frac{2P_r\dot{B}}{B^3}\bigg)+2\mathcal{R}_{22}\bigg(\frac{\dot{P}_\bot}{C^2}-\frac{2P_\bot\dot{C}}{C^3}\bigg)\bigg\}
-\frac{\mu\dot{\mathcal{R}}}{2A^2}\bigg],\\\nonumber
\mathbb{Z}_2&=\frac{2\zeta}{16\pi+\zeta\mathcal{R}}\bigg[\big(P_r\mathcal{R}^{10}\big)^.+\big(P_r\mathcal{R}^{11}\big)'
-\big(\frac{\varsigma
A\mathcal{R}^{00}}{B}\big)^.-\big(\frac{\varsigma
A\mathcal{R}^{11}}{B}\big)'+\dot{\mu}\mathcal{G}^{10}\\\nonumber
&+\mu'\mathcal{G}^{11}-\frac{1}{2B^2}\bigg\{\mathcal{R}_{00}\bigg(\frac{\mu'}{A^2}-\frac{2\mu
A'}{A^3}\bigg)+2\mathcal{R}_{01}\bigg(\frac{\varsigma'}{AB}-\frac{\varsigma
A'}{A}-\frac{\varsigma B'}{B}\bigg)\\\label{A6}
&+\mathcal{R}_{11}\bigg(\frac{P'_r}{B^2}-\frac{2P_rB'}{B^3}\bigg)+2\mathcal{R}_{22}\bigg(\frac{P'_\bot}{C^2}-\frac{2P_\bot
C'}{C^3}\bigg)\bigg\}-\frac{P_r\mathcal{R}'}{2B^2}\bigg].
\end{align}

\section*{Appendix B}

\renewcommand{\theequation}{B\arabic{equation}}
\setcounter{equation}{0} The correction terms of
$f(\mathcal{R},\mathcal{T},\mathcal{R}_{\varphi\vartheta}\mathcal{T}^{\varphi\vartheta})$
theory in scalars \eqref{g28}-\eqref{g31} are
\begin{align}\nonumber
\chi_{1}^{(\mathcal{C})}&=-\frac{8\pi\zeta}{1+\zeta\mu}\bigg[\bigg\{\Box
\mathcal{T}^{\varphi}_{\epsilon}+\frac{1}{2}\nabla_{\vartheta}\nabla^{\varphi}\mathcal{T}^{\vartheta}_{\epsilon}
+\frac{1}{2}\nabla_{\vartheta}\nabla_{\epsilon}\mathcal{T}^{\vartheta\varphi}\bigg\}h^{\epsilon}_{\varphi}
-\mathcal{R}^{\varphi}_{\vartheta}\bigg(P-\frac{\Pi}{3}\bigg)h^{\vartheta}_{\varphi}\\\nonumber
&-\mathcal{R}_{\vartheta\epsilon}\bigg(P-\frac{\Pi}{3}\bigg)h^{\epsilon\vartheta}\bigg]
+\frac{8\pi\zeta}{1+\zeta\mu}\bigg[\bigg\{\mathcal{R}_{\varphi\vartheta}\mathcal{T}^{\varphi\vartheta}
-\frac{1}{2}\nabla_{\varphi}\nabla_{\vartheta}\mathcal{T}^{\varphi\vartheta}\bigg\}
+\nabla_{\vartheta}\nabla_{\varphi}\mathcal{T}^{\vartheta\varphi}\\\label{B1}
&+\frac{1}{2}\Box(\mu-3P)-2\mathcal{R}\bigg(P-\frac{\Pi}{3}\bigg)\bigg],\\\nonumber
\chi_{2}^{(\mathcal{C})}&=\frac{8\pi\zeta}{1+\zeta\mu}\bigg\{\mathcal{R}_{\varphi\vartheta}\mathcal{T}^{\varphi\vartheta}
-\frac{1}{2}\nabla_{\varphi}\nabla_{\vartheta}\mathcal{T}^{\varphi\vartheta}\bigg\}+\frac{4\pi\zeta}{1+\zeta\mu}\bigg[\frac{1}{2}
\big\{\Box\mathcal{T}+\mathcal{K}^{\vartheta}\mathcal{K}^{\delta}\Box\mathcal{T}_{\vartheta\delta}\\\nonumber
&-3\mathcal{K}_{\varphi}\mathcal{K}^{\delta}\Box\mathcal{T}^{\varphi}_{\delta}\big\}
+2\mathcal{R}^{\varphi}_{\vartheta}(Ph^{\vartheta}_{\varphi}+\Pi^{\vartheta}_{\varphi}+\varsigma\mathcal{K}^{\vartheta}
\mathcal{W}_{\varphi})+3\mathcal{R}^{\varphi}_{\vartheta}(\mu\mathcal{K}^{\vartheta}\mathcal{K}_{\varphi}
+\varsigma\mathcal{W}^{\vartheta}\mathcal{K}_{\varphi})\\\nonumber
&+3\mathcal{R}_{\vartheta\delta}(\mu\mathcal{K}^{\vartheta}\mathcal{K}^{\delta}+\varsigma\mathcal{K}^{\delta}\mathcal{W}^{\vartheta})
-\frac{1}{2}\{2\nabla_{\vartheta}\nabla^{\varphi}\mathcal{T}^{\vartheta}_{\varphi}-3\mathcal{K}_{\varphi}\mathcal{K}^{\delta}
\nabla_{\vartheta}\nabla^{\varphi}\mathcal{T}^{\vartheta}_{\delta}\\\nonumber
&-3\mathcal{K}_{\varphi}\mathcal{K}^{\delta}\nabla_{\vartheta}\nabla_{\delta}\mathcal{T}^{\vartheta\varphi}
+\mathcal{K}^{\varphi}\mathcal{K}^{\delta}\nabla_{\vartheta}\nabla_{\varphi}\mathcal{T}^{\vartheta}_{\delta}
+\mathcal{K}^{\varphi}\mathcal{K}^{\delta}\nabla_{\vartheta}\nabla_{\delta}\mathcal{T}^{\vartheta}_{\varphi}\}+\Box(\mu-3P)\\\label{B2}
&-4\mathcal{R}_{\vartheta\epsilon}\mathcal{T}^{\vartheta\epsilon}
+2\nabla_{\vartheta}\nabla_{\epsilon}\mathcal{T}^{\vartheta\epsilon}\bigg],\\\nonumber
\chi_{\varphi\vartheta}^{(\mathcal{C})}&=\frac{2\pi\zeta}{1+\zeta\mu}\bigg[h^{\lambda}_{\varphi}h^{\epsilon}_{\vartheta}\Box
\mathcal{T}_{\lambda\epsilon}-\Box\mathcal{T}_{\varphi\vartheta}-\mathcal{K}_{\varphi}\mathcal{K}_{\vartheta}\mathcal{K}_{\gamma}
\mathcal{K}^{\delta}\Box\mathcal{T}^{\gamma}_{\delta}\bigg]+\frac{4\pi\zeta}{1+\zeta\mu}
\bigg[(h^{\lambda}_{\varphi}\mathcal{R}_{\lambda\delta}\\\nonumber
&-\mathcal{R}_{\varphi\delta})(Ph^{\delta}_{\vartheta}+\Pi^{\delta}_{\vartheta}+\varsigma\mathcal{K}^{\delta}\mathcal{W}_{\vartheta})
+(h^{\epsilon}_{\vartheta}\mathcal{R}_{\delta\epsilon}-\mathcal{R}_{\delta\vartheta})(Ph^{\delta}_{\varphi}+\Pi^{\delta}_{\varphi}
+\varsigma\mathcal{K}^{\delta}\mathcal{W}_{\varphi})\\\nonumber
&-\frac{1}{2}\{h^{\lambda}_{\varphi}h^{\epsilon}_{\vartheta}\nabla_{\delta}\nabla_{\lambda}\mathcal{T}^{\delta}_{\epsilon}
+h^{\lambda}_{\varphi}h^{\epsilon}_{\vartheta}\nabla_{\delta}\nabla_{\epsilon}\mathcal{T}^{\delta}_{\lambda}
-\nabla_{\delta}\nabla_{\varphi}\mathcal{T}^{\delta}_{\vartheta}
-\nabla_{\delta}\nabla_{\vartheta}\mathcal{T}^{\delta}_{\varphi}\\\label{B3}
&-\mathcal{K}_{\varphi}\mathcal{K}_{\vartheta}\mathcal{K}_{\gamma}\mathcal{K}^{\delta}\nabla_{\epsilon}
\nabla^{\gamma}\mathcal{T}^{\epsilon}_{\delta}-\mathcal{K}_{\varphi}\mathcal{K}_{\vartheta}\mathcal{K}_{\gamma}
\mathcal{K}^{\delta}\nabla_{\epsilon}\nabla_{\delta}\mathcal{T}^{\epsilon\gamma}\}\bigg].
\end{align}

The modified corrections in Eqs.\eqref{g51}-\eqref{g53} are
\begin{align}\label{B4}
\chi_1&=\frac{1}{8\pi
B^2}\bigg(\frac{2C''}{C}-\frac{2B'C'}{BC}-\frac{B^2}{C^2}+\frac{C'^2}{C^2}\bigg)+\frac{\dot{C}}{8\pi
C}\bigg(\frac{2\dot{B}}{B}+\frac{\dot{C}}{C}\bigg),\\\label{B5}
\chi_2&=\frac{2\dot{B}\dot{C}}{BC}-\frac{3\ddot{B}}{2B^2}+\frac{\dot{C}^2}{C^2}-\frac{3\ddot{C}}{C}+\frac{\mathcal{R}}{2},\\\label{B6}
\chi_3&=-\frac{2B'C'}{B^3C}+\frac{2B'\dot{C}}{B^3C}+\frac{\dot{B}\dot{C}}{BC}+\frac{\ddot{B}}{2B}+\frac{C'^2}{B^2
C^2}+\frac{C''}{B^2C},\\\label{B7}
\chi_4&=\frac{B'C'}{B^3C}+\frac{\dot{B}\dot{C}}{BC}-\frac{C'^2}{B^2C^2}-\frac{C''}{B^2C}+\frac{\dot{C}^2}{C}
+\frac{\ddot{C}}{C},\\\label{B8}
\chi_5&=\frac{5\dot{B}C'}{BC^2}+\frac{2C'\dot{C}}{BC^2}-\frac{\dot{C}'}{BC},
\quad \chi_6=-\frac{1}{8\pi
B}\bigg(\frac{2\dot{B}C'}{BC}-\frac{2\dot{C}'}{C}\bigg),\\\label{B9}
\chi_7&=\frac{2\dot{C}'}{BC}-\frac{2\dot{B}C'}{B^2C}, \quad
\chi_8=\frac{2C''}{B^2C}-\frac{2B'C'}{B^3C}-\frac{\ddot{B}}{B}-\frac{2\ddot{C}}{C}+\frac{\mathcal{R}}{2},\\\label{B10}
\chi_9&=\frac{1}{8\pi}\bigg(\frac{C'^2}{B^2C^2}-\frac{2\ddot{C}}{C}-\frac{\dot{C}^2}{C^2}-\frac{1}{C^2}\bigg),
\quad
\chi_{10}=\frac{\ddot{B}}{2B}-\frac{\ddot{C}}{C}-\frac{\dot{C}^2}{C^2},\\\label{B11}
\chi_{11}&=\frac{3C''}{B^2C}-\frac{2B'C'}{B^3C}-\frac{2B'\dot{C}}{B^3C}-\frac{3\dot{B}\dot{C}}{BC}-\frac{3
\ddot{B}}{2B}-\frac{C'^2}{B^2C^2}+\frac{\mathcal{R}}{2},\\\label{B12}
\chi_{12}&=\frac{C'^2}{B^2C^2}-\frac{B'C'}{B^3C}-\frac{\dot{B}\dot{C}}{BC}+\frac{C''}{B^2C}-\frac{\dot{C}^2}{C^2}
-\frac{\ddot{C}}{C},\\\label{B13}
\chi_{13}&=\frac{2\dot{C}'}{BC}-\frac{4\dot{B}C'}{B^2C}-\frac{2C'\dot{C}}{BC^2},\\\label{B14}
\chi_{14}&=-\frac{1}{8\pi}\bigg(\frac{B'C'}{B^5C}+\frac{\ddot{B}}{B}-\frac{C''}{B^2C}+\frac{\ddot{C}}{C}\bigg),
\quad
\chi_{15}=-\frac{\dot{B}\dot{C}}{BC}-\frac{\ddot{B}}{2B},\\\label{B15}
\chi_{16}&=\frac{B'C'}{B^3C}-\frac{2B'\dot{C}}{B^3C}-\frac{\dot{B}\dot{C}}{BC}-\frac{\ddot{B}}{2B},\\\label{B16}
\chi_{17}&=\frac{2C''}{B^2C}-\frac{2B'C'}{B^3C}-\frac{2\dot{B}\dot{C}}{BC}+\frac{2C'^2}{B^2C^2}-\frac{2\dot{C}^2}{C^2}-\frac{2
\ddot{C}}{C}-\frac{2}{C^2}+\frac{\mathcal{R}}{2},\\\label{B17}
\chi_{18}&=\frac{B'\dot{B}}{B^3}-\frac{2\dot{B}C'}{B^2C}-\frac{\dot{B}'}{B^2}.
\end{align}


\vspace{0.5cm}

\begin{thebibliography}{43}

\bibitem{1} H.A. Buchdahl, Non-linear Lagrangians and cosmological theory, Mon. Not. R. Astron. Soc. 150 (1970) 1-8.

\bibitem{2} S. Nojiri, S.D. Odintsov, Modified gravity with negative and positive powers of curvature:
Unification of inflation and cosmic acceleration, Phys. Rev. D 68
(2003) 123512.

\bibitem{2a} G. Cognola, E. Elizalde, S. Nojiri, S.D. Odintsov, S. Zerbini, One-loop $f(R)$ gravity in de Sitter universe, J. Cosmol. Astropart. Phys.
2005 (2005) 010.

\bibitem{2b} Y.S. Song, W. Hu, I. Sawicki, Large scale structure of $f(R)$ gravity, Phys.
Rev. D 75 (2007) 044004.

\bibitem{2d} M. Sharif, Z. Yousaf, Stability of the charged spherical dissipative collapse in $f(R)$ gravity, Mon. Not. R. Astron. Soc. 434 (2013) 2529-2538.

\bibitem{10} O. Bertolami, C.G. Boehmer, T. Harko, F.S.N. Lobo, Extra force in $f(R)$ modified theories of gravity, Phys. Rev. D 75 (2007) 104016.

\bibitem{20} T. Harko, F.S.N. Lobo, S. Nojiri, S.D. Odintsov, $f(R,T)$ gravity, Phys. Rev. D 84 (2011) 024020.

\bibitem{21} M. Sharif, M. Zubair, Thermodynamic behavior of particular $f(R,T)$-gravity models, J. Exp. Theor. Phys. 117 (2013) 248-257.

\bibitem{21a} H. Shabani, M. Farhoudi, $f(R,T)$ cosmological models in phase space, Phys. Rev. D 88 (2013) 044048.

\bibitem{21e} M. Sharif, A. Siddiqa, Study of charged stellar structures in $f(R,T)$ gravity, Eur. Phys. J. Plus 132 (2017) 1-10.

\bibitem{21f} A. Das, S. Ghosh, B.K. Guha, S. Das, F. Rahaman, S. Ray, Gravastars in $f(R,T)$ gravity, Phys. Rev. D 95 (2017) 124011.

\bibitem{21g} M. Zubair, H. Azmat, I. Noureen, Dynamical analysis of cylindrically symmetric anisotropic sources in $f(R,T)$ gravity,
Eur. Phys. J. C 77 (2017) 169.

\bibitem{22} Z. Haghani, T. Harko, F.S.N. Lobo, H.R. Sepangi, S. Shahidi, Further matters in space-time geometry:
$f(R,T,R_{\mu\nu} T^{\mu\nu})$ gravity, Phys. Rev. D 88 (2013)
044023.

\bibitem{22a} M. Sharif, M. Zubair, Study of thermodynamic laws in $f(R,T,R_{\mu\nu} T^{\mu\nu})$ gravity,
J. Cosmol. Astropart. Phys. 2013 (2013) 042.

\bibitem{22b} M. Sharif, M. Zubair, Energy conditions in $f(R,T,R_{\mu\nu} T^{\mu\nu})$ gravity, J. High Energy Phys. 2013 (2013) 1-21.

\bibitem{23} S.D. Odintsov, D. S{\'a}ez-G{\'o}mez, $f(R,T,R_{\mu\nu} T^{\mu\nu})$ gravity phenomenology and $\Lambda$CDM universe,
Phys. Lett. B 725 (2013) 437-444.

\bibitem{25} E.H. Baffou, M.J.S. Houndjo, J. Tosssa, Exploring stable models in $f(R,T,R_{\mu\nu} T^{\mu\nu})$ gravity,
Astrophys. Space Sci. 361 (2016) 1-8.

\bibitem{25a} M. Sharif, A. Waseem, Physical behavior of anisotropic compact stars in $f(R,T,R_{\mu\nu} T^{\mu\nu})$ gravity, Can. J. Phys. 94 (2016) 1024-1039.

\bibitem{26} Z. Yousaf, M.Z. Bhatti, T. Naseer, Study of static charged spherical structure in $f(R,T,Q)$ gravity, Eur. Phys. J. Plus
135 (2020) 1-21.

\bibitem{26a} Z. Yousaf, M.Z. Bhatti, T. Naseer, Evolution of the charged dynamical radiating spherical structures, Ann. Phys.
420 (2020) 168267.

\bibitem{26b} Z. Yousaf, M.Z. Bhatti, T. Naseer, Measure of complexity for dynamical self-gravitating structures, Int. J. Mod. Phys. D 29 (2020) 2050061.

\bibitem{26c} Z. Yousaf, M.Y. Khlopov, M.Z. Bhatti, T. Naseer, Influence of modification of gravity on the complexity factor of static spherical structures,
Mon. Not. R. Astron. Soc. 495 (2020) 4334-4346.

\bibitem{26d} Z. Yousaf, M.Z. Bhatti, T. Naseer, New definition of complexity factor in $f(R,T,R_{\mu\nu} T^{\mu\nu})$ gravity, Phys. Dark Universe
28 (2020) 100535.

\bibitem{26e} Z. Yousaf, M.Z. Bhatti, T. Naseer, I. Ahmad, The measure of complexity in charged celestial bodies in $f(R,T,R_{\mu\nu} T^{\mu\nu})$ gravity,
Phys. Dark Universe 29 (2020) 100581.

\bibitem{27} M. Sharif, T. Naseer, Effects of $f(R,T,R_{\gamma\upsilon}T^{\gamma\upsilon})$ gravity on anisotropic charged compact structures,
Chin. J. Phys. 73 (2021) 179-194.

\bibitem{27a} T. Naseer, M. Sharif, Study of decoupled anisotropic solutions in $f(R,T,R_{\rho\eta}T^{\rho\eta})$ theory,
Universe 8 (2022) 62.

\bibitem{27b} M. Sharif, T. Naseer, Effects of non-minimal matter-geometry coupling on embedding class-one anisotropic solutions,
Phys. Scr. 97 (2022) 055004.

\bibitem{27c} M. Sharif, T. Naseer, Influence of charge on extended decoupled anisotropic solutions in $f(R,T,R_{\lambda\xi}T^{\lambda\xi})$
gravity, arXiv:2203.03268v1 [gr-qc] (2022).

\bibitem{28} R. Lopez Ruiz, H.L. Mancini, X. Calbet, A statistical measure of complexity, Phys. Lett. A 209 (1995) 321.

\bibitem{29} X. Calbet, R. Lopez Ruiz, Tendency towards maximum complexity in a nonequilibrium isolated system, Phys. Rev. E 63 (2001) 066116.

\bibitem{29a} R.G. Catalan, J. Garay, R. Lopez Ruiz, Features of the extension of a statistical measure of complexity to continuous systems, Phys. Rev. E 66 (2002)
011102.

\bibitem{30} J. Sanudo, R. Lopez Ruiz, Statistical complexity and Fisher Shannon information in the H atom, Phys. Lett. A 372 (2008) 5283.

\bibitem{30a} J. Sanudo, A.F. Pacheco, Complexity and white-dwarf structure, Phys. Lett. A 373 (2009) 807.

\bibitem{31} L. Herrera, New definition of complexity for self-gravitating fluid distributions: The spherically symmetric, static case, Phys. Rev. D 97 (2018) 044010.

\bibitem{32} M. Sharif, I.I. Butt, Complexity factor for charged spherical system, Eur. Phys. J. C 78 (2018) 688.

\bibitem{32a} M. Sharif, I.I. Butt, Complexity factor for static cylindrical system, Eur. Phys. J. C 78 (2018) 850.

\bibitem{33} L. Herrera, A. Di Prisco, J. Ospino, Definition of complexity for dynamical spherically symmetric dissipative self-gravitating fluid distributions, Phys. Rev.
D 98 (2018) 104059.

\bibitem{34} L. Herrera, A. Di Prisco, J. Ospino, Complexity factors for axially symmetric static sources, Phys. Rev. D 99 (2019) 044049.

\bibitem{35} M. Sharif, A. Majid, Complexity factor for static sphere in self-interacting Brans-Dicke gravity, Chin. J. Phys. 61 (2019) 38.

\bibitem{35a} M. Sharif, A. Majid, Complexity of dynamical sphere in self-interacting Brans-Dicke gravity, Eur. Phys. J. C 80 (2020) 1185.

\bibitem{35c} M. Sharif, A. Majid, Complexity factor for cylindrical system in Brans-Dicke gravity, Indian J. Phys. 95 (2021) 769-777.

\bibitem{36a} M. Zubair, H. Azmat, Complexity analysis of dynamical spherically-symmetric dissipative self-gravitating objects in
modified gravity, Int. J. Mod. Phys. D 29 (2020) 2050014.

\bibitem{36b} M. Zubair, H. Azmat, Complexity analysis of cylindrically symmetric self-gravitating dynamical system in $f(R,T)$ theory of gravity, Phys. Dark Universe
28 (2020) 100531.

\bibitem{36d} M. Sharif, K. Hassan, Complexity for dynamical anisotropic sphere in $f(G,T)$ gravity, Chin. J. Phys. 77 (2022) 1479-1492.

\bibitem{36e} M. Sharif, K. Hassan, Complexity of dynamical cylindrical system in $f(G,T)$ gravity, Mod. Phys. Lett. A 37 (2022) 2250027.

\bibitem{39} Y.Y. Zhao, Y.B. Wu, J.B. Lu, Z. Zhang, W.L. Han, L.L. Lin, Modified $f(G)$ gravity models with curvature-matter coupling,
Eur. Phys. J. C 72 (2012) 1-10.

\bibitem{40} H. Yu, W.D. Guo, K. Yang, Y.X. Liu, Scalar particle production in a simple Horndeski theory, Phys. Rev. D 97 (2018) 083524.

\bibitem{41ba} K.S. Thorne, Absolute stability of Melvin's magnetic universe, Phys. Rev. 139 (1965) B244.

\bibitem{41bb} L. Herrera, J. Ospino, A. Di Prisco, E. Fuenmayor, O. Troconis, Structure and evolution of self-gravitating objects and the
orthogonal splitting of the Riemann tensor, Phys. Rev. D. 79 (2009)
064025.

\bibitem{42bd} C.J. Hansen, S.D. Kawaler, V. Trimble, Stellar Interiors: Physical Principles, Structure and Evolution, Springer, New York, 1994.
\end{thebibliography}
\end{document}